\documentclass{article}

\usepackage[utf8]{inputenc} 
\usepackage[T1]{fontenc}    
\usepackage{hyperref}       
\usepackage{url}            
\usepackage{booktabs}       
\usepackage{amsfonts}       
\usepackage{nicefrac}       
\usepackage{microtype}      
\usepackage{lipsum}
\usepackage{float}
\usepackage{booktabs}
\usepackage[numbers]{natbib}
\usepackage{amsthm, amsmath, mathtools, amssymb}

\usepackage[utf8]{inputenc} 
\usepackage{graphicx}
\usepackage{xcolor}
\usepackage{epsfig}
\usepackage{algpseudocode}
\usepackage{algorithm}
\usepackage[top=2.5cm,bottom=2.5cm,left=3.5cm,right=3.5cm]{geometry}
\usepackage{caption}
\usepackage{subcaption}

\begin{document}

\title{Where Are We? \\ Using Scopus to Map the Literature at the Intersection Between Artificial Intelligence and Research on Crime}
\author{Gian Maria Campedelli\thanks{} \\
  \footnotesize  University of Trento\\
  \footnotesize  Trento, Italy \\
  \footnotesize \texttt{gianmaria.campedelli@unitn.it}}

\maketitle

\begin{abstract}
Research on Artificial Intelligence (AI) applications has spread over many scientific disciplines. Scientists have tested the power of intelligent algorithms developed to predict (or learn from) natural, physical and social phenomena. This also applies to crime-related research problems. Nonetheless, studies that map the current state of the art at the intersection between AI and crime are lacking. What are the current research trends in terms of topics in this area? What is the structure of scientific collaboration when considering works investigating criminal issues using machine learning, deep learning, and AI in general? What are the most active countries in this specific scientific sphere? Using data retrieved from the Scopus database, this work quantitatively analyzes 692 published works at the intersection between AI and crime employing network science to respond to these questions. Results show that researchers are mainly focusing on cyber-related criminal topics and that relevant themes such as algorithmic discrimination, fairness, and ethics are considerably overlooked. Furthermore, data highlight the extremely disconnected structure of co-authorship networks. Such disconnectedness may represent a substantial obstacle to a more solid community of scientists interested in these topics. Additionally, the graph of scientific collaboration indicates that countries that are more prone to engage in international partnerships are generally less central in the network. This means that scholars working in highly productive countries (e.g. the United States, China) tend to mostly collaborate domestically. Finally, current issues and future developments within this scientific area are also discussed.
\end{abstract}


\section{Introduction}
\label{intro}
The last two decades have witnessed a growing interest of scholars coming from natural, physical, and mathematical sciences in social science problems. Mathematical and statistical modeling have widespread across multiple disciplines that focus on the study of human beings and societies and that have traditionally been marked by qualitative research. Besides economics, which inherently deal with numerical quantities and is, therefore, traditionally more receptive in adopting quantitative approaches, mathematics and statistics have infiltrated many other disciplines falling under the broad category of ``social sciences'', including sociology, political science, and criminology \cite{AxelrodAdvancingArtSimulation1997, GoldthorpeCausationStatisticsSociology2001, SawyerArtificialSocietiesMultiagent2003, BorgattiNetworkAnalysisSocial2009b, PiqueroHandbookQuantitativeCriminology2010, NaldiMathematicalModelingCollective2010, SubrahmanianHandbookComputationalApproaches2013b, JohnsonPoliticalScienceResearch2019}. \\
While the wall of resistance against quantitative research in the social sciences was finally collapsing, opening new perspectives and posing new challenges to scientific inquiry, other fields were experiencing another revolution, potentially one of the most intriguing and fascinating in human history. The interplay between neuroscience, computer science, mathematics, and other satellite fields had, in fact, given light to decisive progress in the formalization, development, and deployment of intelligent algorithms for solving different classes of problems \cite{NilssonQuestArtificialIntelligence2009, RussellArtificialIntelligenceModern2010a}. Artificial Intelligence, through several approaches and hundreds of different algorithms, has since then increasingly become a central component of research in computers and computation and has acquired a critical role in several other fields. The capabilities of AI systems have been tested also in social science fields. \\
Even in this case, such a process has started to contaminate the study of crime. Nevertheless, studies that investigate the extent to which AI has intersected research on crime do not exist. Despite the relevant debates that have emerged regarding two areas of application of AI systems, namely criminal justice and policing \citep{ShapiroReformpredictivepolicing2017, BerkFairnessCriminalJustice2018}, the literature lacks a mapping of the research production that integrate intelligent algorithms and the analysis of offenders and criminal behaviors. In light of these considerations, \color{black} this work proposes to map the extant literature using Scopus, a database containing over 69 million abstract and citation records of peer-reviewed literature.\\ \color{black}
The aim is to shed light on existing trends and patterns in this growing and heterogeneous area of research and to reason about future likely pathways and directions. \\
The article outlines as follows. Section \ref{sec:back} (``Background'') will briefly portray the diffusion of AI applications in social science, also outlining the issues associated with the deployment of intelligent systems in criminal justice and policing. Following, Section \ref{sec:strat} (``Analytic Strategy'') will describe the search strategy and the methodological setup of the study. Section \ref{sec:ana} (``Analysis and Results'') will then present the outcome of the two different analytical dimensions of the study, namely the analysis of current patterns in topics and themes of research related to AI and crime and the structure of individual- and country-level collaboration networks. Finally, in Section \ref{sec:disc} (``Where To, Now? Discussion and Future Developments''), considerations derived from the analyses will be drawn in the attempt to better picture this strand of research and to define its current issues and potential future pathways.

\section{Background}
\label{sec:back}

The study of crime is an area of scientific inquiry that has long benefited from the dialogue among different fields. Crimes and criminal behaviors have been studied from a manifold of perspectives during the last two centuries. Criminology itself has been enriched by intersections and debates across disciplines such as medicine, psychology, biology, philosophy, law, sociology, economics, and political science. Despite the different scientific trajectories of each of the separate fields sharing the interest about crime, one common process applies to many of them: the increasing use of data to propose, test or support theories and, more broadly, the growing prevalence of quantitative research \cite{TewksburyMethodologicalorientationsarticles2005, SampsonGoldStandardMyths2010}.

This process has been certainly favored by technological and scientific signs of progress made in the last fifty years in other scientific fields (e.g., the diffusion of personal computers), and has been facilitated by the interest of policy- and decision-makers in designing criminal strategies and counter-policies based on empirical evidence. Regardless of the specific topic being investigated, quantitative and statistical methods have gained success and fostered the interest of scholars that do not belong to the founding fields of criminology. Hence, the rapid availability of data and information for measuring, mapping, explaining, predicting, and forecasting crime has mathematicians, statisticians, physicists, and computer scientists to considerably contribute to the study of crime.

Nonetheless, while the quantitative shift in the study of crime is irrefutable, researchers have not yet scanned the scientific production that employs artificial intelligence (AI) to investigate crime-related problems. In the last ten years, due to the combined effect of several events and phenomena related to the study of artificial intelligence and statistical learning, the success of algorithms designed to learn existing patterns in data without being explicitly programmed to do so has been enormous. Artificial intelligence relies on legacies of mathematical constructions and techniques that are centuries old \citep{RussellArtificialIntelligenceModern2010a}. However, in the last fifteen years, especially due to the breakthroughs in the use of neural networks, artificial intelligence has gained unprecedented attention and popularity beyond the borders of academia.
The AI landscape in terms of approaches and methods is extremely complex and continuously evolving. Nonetheless, expressions such as ``machine learning'' and ``deep learning'' have become popular also to non-specialists and non-academics. 

This wide success has led to a displacement of debates, applications, and experiments in areas other than computer science and mathematics. This displacement has indeed touched social sciences or, more narrowly, specific societal problems. To exemplify, methods based on machine and deep learning have been used to predict poverty using a variety of data sources, including satellite images \cite{XieTransferLearningDeep2016a, JeanCombiningsatelliteimagery2016a}. The availability of rich and multi-modal data and the strengths of intelligent algorithms have made also possible to study topics related to climate change and models \cite{TripathiDownscalingprecipitationclimate2006, GangulyDataMiningClimate2008, RaspDeeplearningrepresent2018}. Additional applications have focused on social work settings as well, proposing strategies and models relying on AI to minimize violence in homeless youth or to ameliorate the living conditions of homeless people \cite{TambeArtificialIntelligenceSocial2018}. Other relevant applications have focused on agricultural issues \cite{KussulDeepLearningClassification2017, KamilarisDeeplearningagriculture2018}, health care \cite{ChenScalableApproachesHome2017, MohamedShakeelMaintainingSecurityPrivacy2018, MiottoDeeplearninghealthcare2018}, traffic prediction and transport optimization \cite{LvTrafficFlowPrediction2014, HuangDeepArchitectureTraffic2014} and individual and collective behaviors on social media \cite{LiuEarlyDetectionFake2018, DiMininMachinelearningtracking2018, YangArmingpublicartificial2019}.

With different level of sophistication and performance, studies have also addressed crime-related problems \citep{BerkMachineLearningRisk2019,WangDeepLearningRealTime2017a,SavageDetectionMoneyLaundering2017, HuangDeepCrimeAttentiveHierarchical2018a}.

Beyond studies focusing on their mere application, however, AI systems have triggered conceptual, moral, and philosophical debates on ethics, fairness, and accountability, given the increasing number of real-world settings in which algorithmic decision-making is deployed every day \citep{EtzioniIncorporatingEthicsArtificial2017}. These debates have been centered around the idea of ensuring the use of these technological and scientific advances for ensuring social good, aiming at sensitizing the scientific community and the civil society on the concrete harms that minorities and other disadvantaged strata of the population face due to these systems.

There exist several realms in which the use of intelligent algorithms has raised different types of concerns, in terms of ethics, respect of human rights, and political impacts \citep{LinRobotEthics2017, EtzioniIncorporatingEthicsArtificial2017, YeungAlgorithmicregulationcritical2018, HelbingWillDemocracySurvive2019}. In parallel with the vivid debate on themes such as superintelligence and existential risk \citep{MullerFutureProgressArtificial2016}, research groups, policy-makers, and activists are pushing towards the definition of guidelines and the improvements of current practices to make AI safer and more ethical \citep{OsobaIntelligenceOurImage2017, LepriFairTransparentAccountable2018, GebruRaceGender2020, Hannacriticalracemethodology2020}. Autonomous vehicles, face recognition tools, data privacy, biometrics are currently some of the hot points of the discussion on the pitfalls of unregulated or poorly regulated AI. Criminal justice and policing, two critical dimensions of research in criminology, are also part of this animated debate, given the diffused deployment of criminal justice risk assessment tools \citep{BerkMachineLearningRisk2019} and predictive policing software to support law enforcement activities \citep{PerryPredictivePolicingRole2013}. 
Notwithstanding the very recent formation of this area of scholarly investigation, research at the intersection between AI and crime is developing not only in the direction of practical applications, use-cases, and targeted experiments but also concerning the mobilization of the scientific community towards a fair and non-discriminatory use of AI systems in the real-world. This two-fold process is likely to converge in the next years (beyond being an expectation, this is indeed a hope), and might help to create a highly complex and diverse research community marked by heterogeneous backgrounds and a multitude of specific interests. 

To better understand the dynamics of this scientific area and portray the current scenario, however, studies are required that focus on trends, topics and collaboration structures. This study seeks to contribute in this direction.

\section{Analytic Strategy}
\label{sec:strat}
\subsection{Search Strategy and Methods}
In order to gather the data needed to map the existing literature that applies artificial intelligence in the attempt to study crime (in a broad sense), I have performed a search on the Scopus database. Scopus contains over 69 million abstract and citation records of peer-reviewed literature in a wide variety of disciplines. After multiple tests, the chosen query has been (CRIM$^{*}$ or CRIMINAL$^{*}$ or CRIMINOLOG$^{*}$) AND (``MACHINE LEARNING'' OR ``DEEP LEARNING'' OR ``ARTIFICIAL INTELLIGENCE'').\footnote{The search was performed on November 14th, 2019.}
The query has been kept sufficiently broad to avoid the exclusion of relevant records from the search. The working assumption is that publications at the intersection between AI and crime, though specifically directed to particular types of criminal phenomena or methodological approaches, are highly likely to mention general terms such as ``crime/s", ``criminogenic'', ``criminal/s``, ``criminality'', ``criminalization", ``criminology'',  ``criminological'' for the crime-related set part, and at least one expression among ``machine learning'', ``deep learning'' and ``artificial intelligence''.

Tests with longer and more complex queries, e.g., queries listing different types of crimes or different types of algorithmic approaches, provided fewer results than the general query, given that, although sparse, the field encompasses a great variety of approaches and crime-related problems. For this reason, this general query has been selected as the most appropriate for the aims of the work. Furthermore, I have tested it in two different fields, namely ``TITLE-ABS-KEY'' and ``ABS''. The ``TITLE-ABS-KEY'' searches the desired word in title, abstract, and keywords. Keyword themselves combine different subfields, namely ``AUTHKEY'', ``INDEXTERMS'', ``TRADENAME'' and ``CHEMNAME''. The ``ABS'' field, instead, searches the requested words or expressions in the abstract alone. A first test using the ``TITLE-ABS-KEY'' fields has retrieved a total of 5,161 records. Unfortunately, a random search across the obtained records showed that there was a considerable share of false-positive items. Analyzing such false positive items (namely items that do not deal at all with crime-related topics), I have found that such false positives were driven by errors in the Index Keywords (```INDEXTERMS'').

The Index Keywords are different from Author Keywords (``AUTHKEY'') because they are not provided by the authors. They are, instead, manually added by a team of Scopus professional indexers based on several vocabularies, as the Ei Thesaurus for engineering, technology, and physical sciences, or MeSH for life and health sciences. Errors in false positives, for instance, included articles that were focusing on computer vision techniques to avoid the corruption of images. The terms corruption, in those specific cases, was wrongly intended as related to the crime of corruption, and therefore the indexed keyword ``Crime'' was also added to the list. Given the number of such false positives, I have performed the search only scanning the presence of the queried terms in the abstracts. The search finally retrieved 692 items.

In terms of methods, besides first descriptive statistics regarding the temporal distribution of the publications in the sample and the comparison with works covering AI topics and applications in general, two different analytical dimensions will be investigated. These dimensions respectively aim at (1) investigating patterns of themes and topics in terms of author and index keywords and (2) studying the structure of co-authorship and country-level collaboration of the considered works. Both aims will be pursued by applying network science as the methodological framework. Network science has proven to be an extremely promising scientific field. Derived from mathematical graph theory, it nowadays encompasses many areas including social networks, biological networks, transportation networks, and communication networks. Among the many areas in which network science has shown its potential, stands the field called ``science of science''. Science of science is the quantitative study of how scientific agents (e.g., authors, universities) interact, focusing on the pathways that lead to scientific discovery and aims at better understanding what drives successful contributions \citep{FortunatoSciencescience2018}.

Given the perfectly fitting nature of networks in capturing relations between entities, network science has, therefore, become a mainstream approach to unfold the characteristics and patterns across scientific domains. For instance, networks have been useful in studying co-authorship in management and organizational studies \citep{AcedoCoAuthorshipManagementOrganizational2006}, the structure of regional innovation system research \citep{LeeInvestigatingstructureregional2010}, scientific endorsement \citep{DingScientificcollaborationendorsement2011}, trends in creativity research \citep{ZhangKnowledgemapcreativity2015}, the characteristics of research community and their evolution over time \citep{LeoneSciabolazzaDetectinganalyzingresearch2017}. 

In light of the success gained by networks in studying how scientists behave and how science occurs, this paper will employ graphs to address the abovementioned aims.

\subsection{Limitations}
There are two layers of limitations for this approach. The first one is inherently related to the fact that Scopus is not the only available database for electronic records of peer-review literature. While Scopus has been used extensively in the literature to survey or map a variety of specific scientific areas \citep{KarpagamMappingnanosciencenanotechnology2011, BornmannMappingexcellencegeography2011, BarelloPatientEngagementEmerging2012, NataleMappingresearchaquaculture2012, RodriguesMappingpatientsafety2014, DevilleMappingCeramicsResearch2015, FahimniaGreensupplychain2015} Web of Science represents a valid alternative. \color{black} There is, in fact, an entire area of research devoted to comparing the two \citep{GavelWebScienceScopus2008, MongeonjournalcoverageWeb2016}. I have tested the extent to which titles overlap in Scopus and Web of Science and, while the total number of gathered items was very similar, there was a far lower similarity in terms of overlap. I have thus checked the nature of the items that were retrieved through the Web of Science search but not through the Scopus one, and vice-versa. Almost 60\% of the items in the Web of Science pool was also present in Scopus. Instead, only about 50\% of the works in the Web of Science pool was present in the Scopus one. Comparing the non-overlapping items, it appeared that the ones excluded by the Web of Science search were far more relevant for the present research compared to those excluded by the Scopus search. Furthermore, another discriminant feature that led to choosing Scopus over Web of Science was the generation procedure for Index keywords and Keywords Plus. While, as already described, the former ones are inserted manually by professional indexers based on the content of each item, the latter are automatically generated based on the titles of the cited works for each retrieved item. Comparing the two, Scopus index keywords were found to be on average much higher than Web of Science Keyword Plus (in terms of frequency) and much more informative, generating a richer pool of information to rely on. In any event, although Scopus appeared to be a better choice for conducting this first assessment of the research at the intersection between AI and crime, the plans for future follow-up works in the next years will have to consider Web of Science as a relevant source of information and, ideally, provide an integration of the two to comprehensively scan this research landscape. \color{black}  

The second limitation regards the decision to search the desired key-expressions in abstracts alone. There is a certain probability that articles that focus on AI applications for crime-related problems do not mention at least one of the expressions included in the two sets of information in their abstracts. In this case, excluding the keywords (both author and indexed ones) from the search, these records would be excluded from the data gathering. 
In summary, the reader shall keep in mind that the results presented in this work are not to be intended as universal, given that the search certainly does not provide the entire universe of publications at the intersection between AI and crime. Nonetheless, given that Scopus is one of the largest databases of scientific literature and that the query is sufficiently broad to guarantee to avoid the exclusion of relevant sources, the results of the study are solid enough for the purposes of the present study.

\section{Analysis and Results}
\label{sec:ana}

\subsection{Data Overview}
\begin{figure}[!htb]
    \centering
    \includegraphics[scale=0.37]{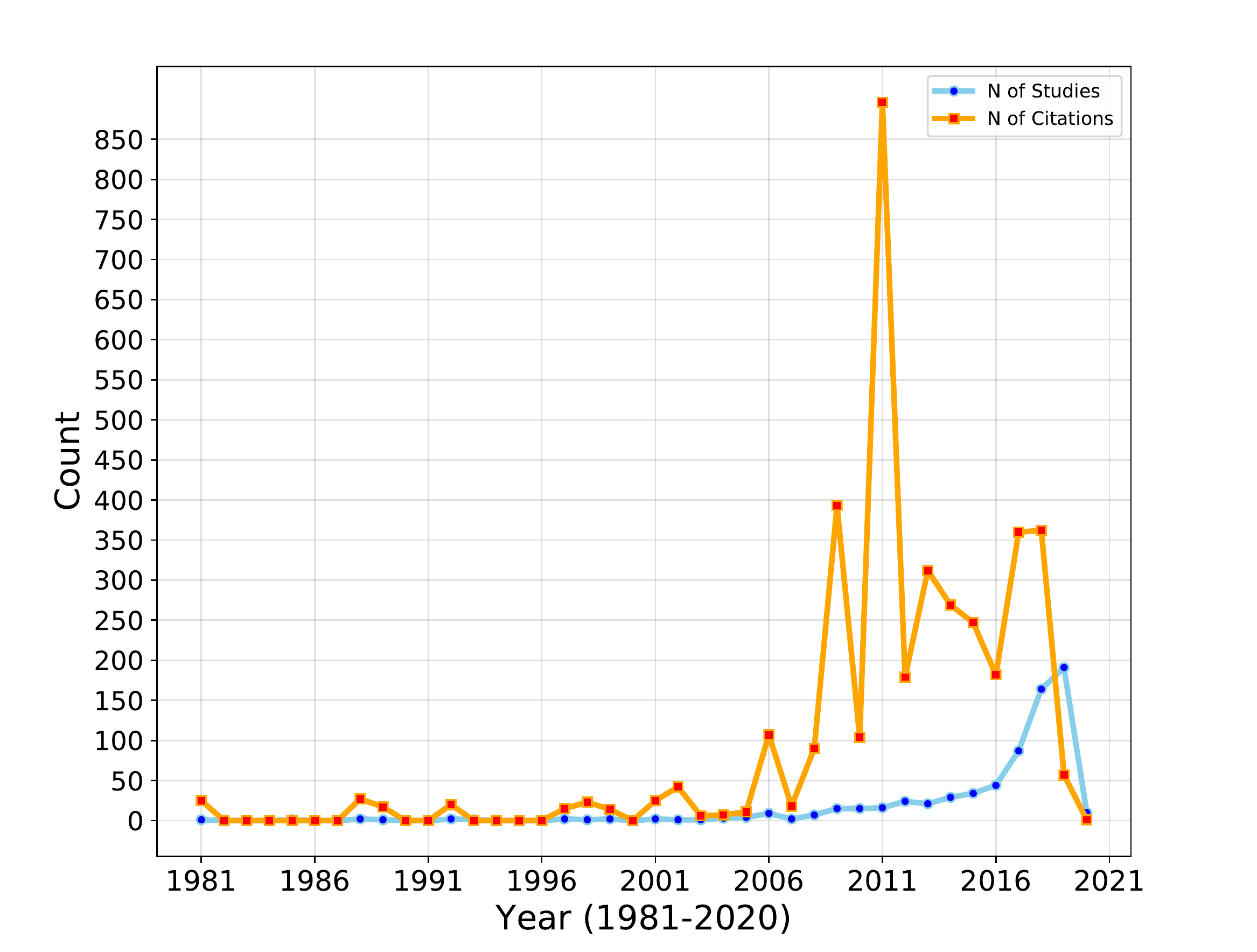}
    \caption{Yearly Trends for Number of Publications and Number of Citations}
    \label{fig:couns}
\end{figure}

In total, 692 studies have been retrieved through the abovementioned query. The export options of Scopus allow obtaining a variety of information on each study, ranging from the year in which it was published to the funding institution. Figure \ref{fig:couns} demonstrates a sensible increase in the number of studies that are published every year, especially in the last five years. The trend in terms of citations is less clear, as its variance is higher but overall shows an increasing behavior as well. 
\newpage

\begin{figure}[!hbt]
    \centering
    \includegraphics[scale=0.41]{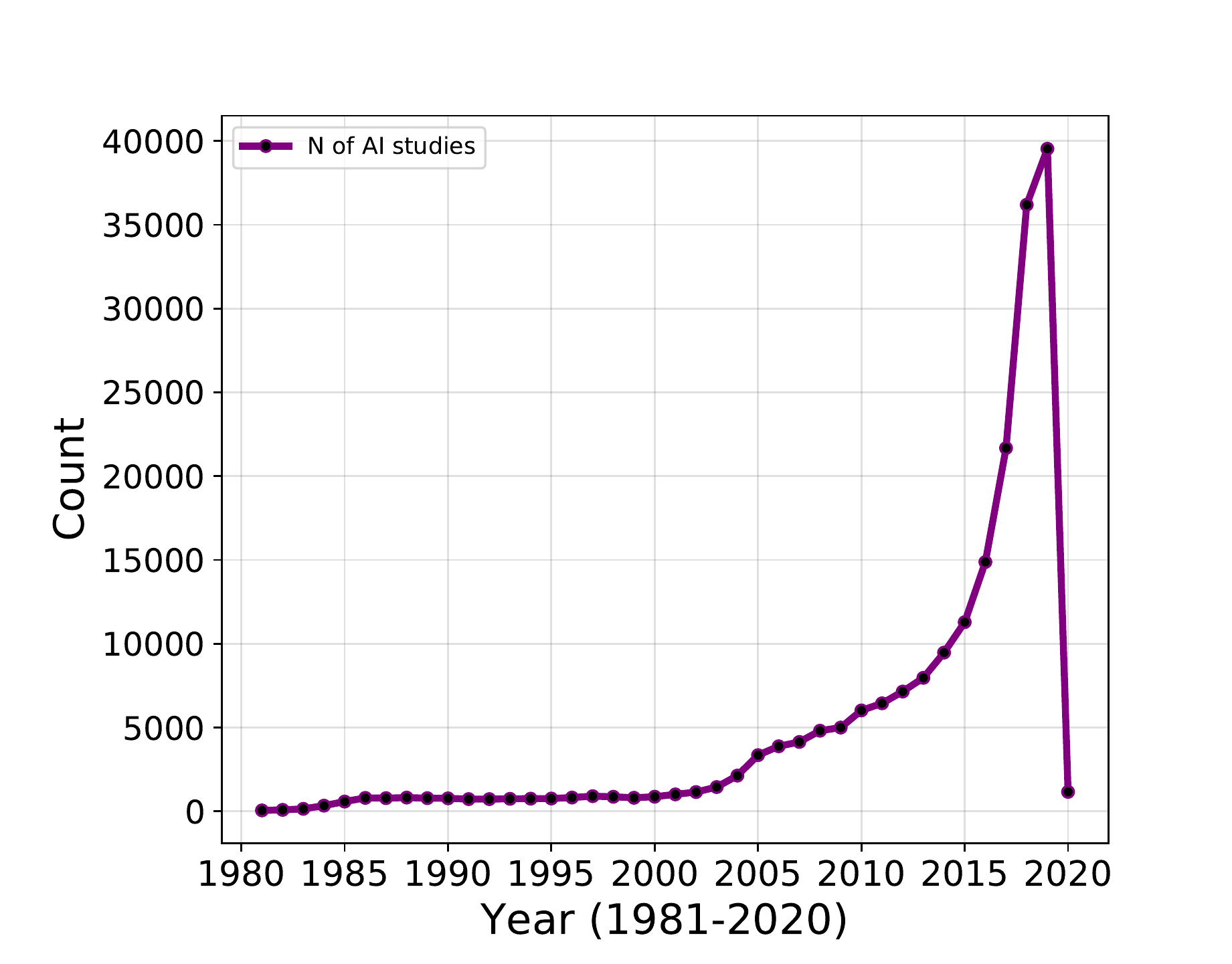}
    \caption{Yearly Number of Publications on AI - Overall}
    \label{fig:ai_overall}
\end{figure}

\begin{figure}[!hbt]
    \centering
    \includegraphics[scale=0.41]{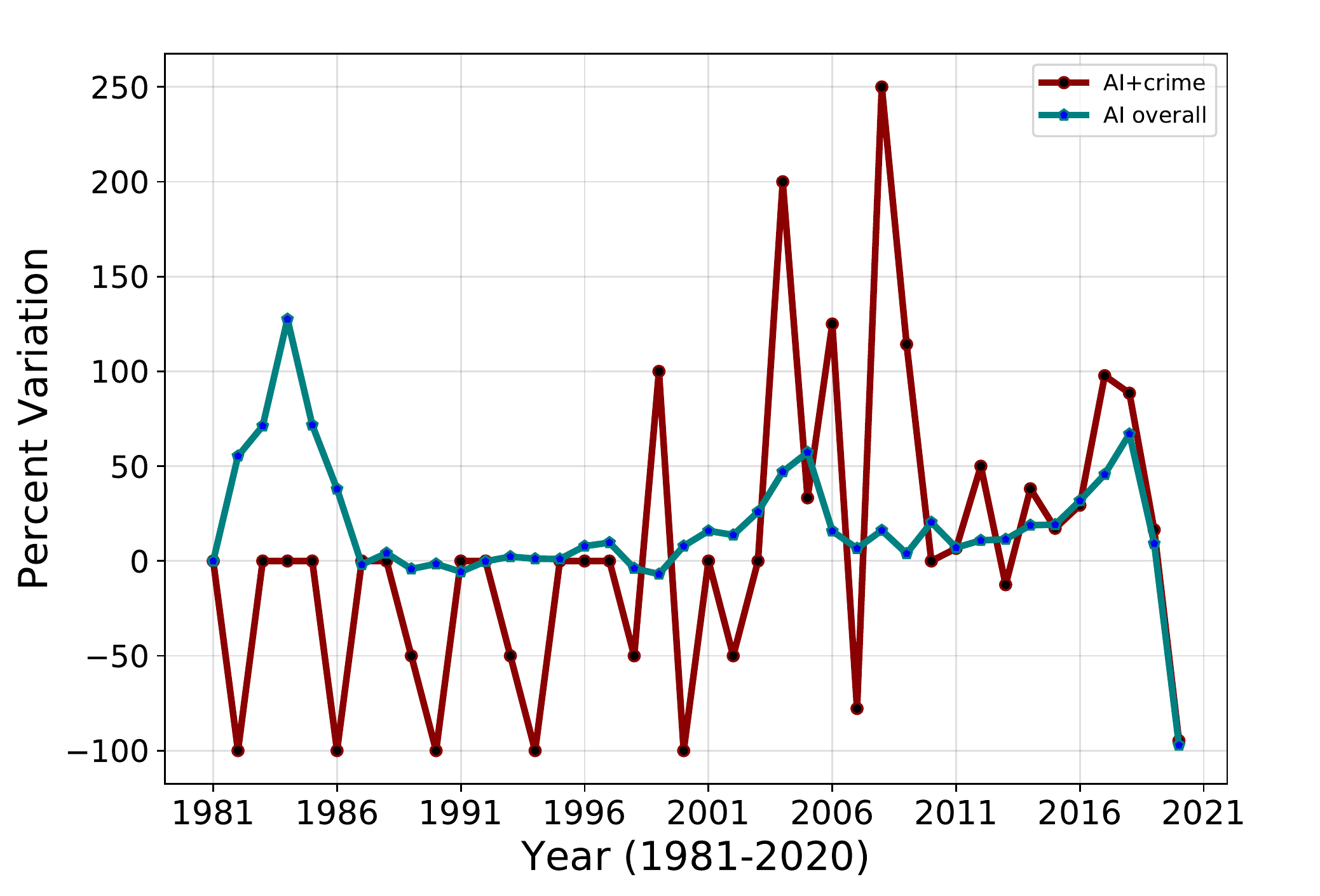}
    \caption{Comparison of Yearly Percent Variations of AI+crime and Overall AI Studies}
    \label{fig:variations}
\end{figure}

It is interesting to compare the trends of yearly publications with the overall trend of works dealing with AI (both at the theoretical and applied levels). For this reason, I have performed a search in Scopus excluding the first part of the query (i.e., excluding crime-related expressions), and considering the same time-frame (namely 1981-2020). The count of studies in Figure \ref{fig:ai_overall} shows that the trend is steeply growing in the last 15 years (monotonically in the last 10 years, with the only exception of 2020 which only includes early publications). However, the plot of percent variations that compares the temporal trends of works at the intersection of AI and crime and overall AI publications better captures the yearly differences between the two (Figure \ref{fig:variations}).  On one hand, the overall AI trend highlights the historical patterns of global research on AI in the late '80s and '90s, where the so-called ``AI winter'' \cite{RussellArtificialIntelligenceModern2010a}, a period of reduction in terms of funding and interest in AI worldwide, hit research. The number of publications started again to increase after 1996 for AI and reached important positive peaks in variation in the years 2003, 2004, 2005. From those years, overall research on AI has continuously increased in the number of publications, reaching a maximum of +66.98\% yearly variation between 2018 and 2019. 

On the other hand, the publications at the intersection between AI and crime were extremely rare and sparsely distributed during the first 20 years. This is probably because research in AI was still confined to a restricted number of scientific and academic fields. To this, it should be added that the fluctuating fortunes of AI in those decades have certainly impacted its diffusion to other areas. After 2000, the number of works has started to sensibly increase. The variations became much more intense and generally positive, except for 2007 (-\%77.77). Notably, in the last three years (2020 excluded), the percent variations of works at the intersection between AI and crime were positive and higher than those for overall AI works (2017: +97.72\% against +45.68\%; 2018: +88.50\% vs +66.98\%; 2019: +16.46\% vs +9.22\%). These figures clearly point in the direction of a growing interest in AI application in the realm of crime-related research problems.


\begin{figure}[!hbt]
    \centering
    \includegraphics[scale=0.5]{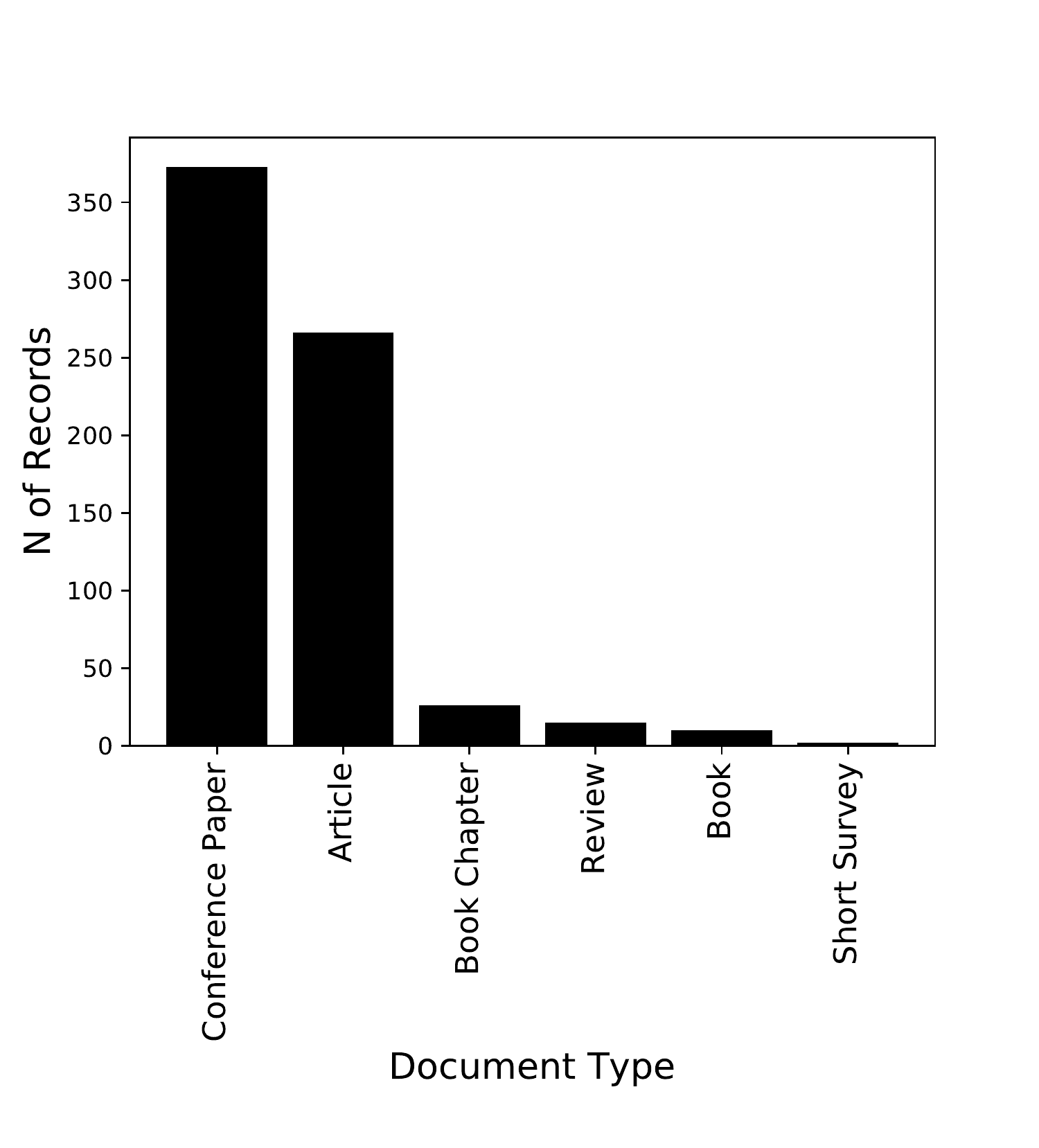}
    \caption{Number of Records per Document Type}
    \label{fig:my_label}
\end{figure}

When focusing on the types of documents obtained from the search (Figure \ref{fig:my_label}), it is interesting to note that the majority of records are related to conference papers (373 against 266 journal articles). This might be due to two factors. First, publishing articles that propose new methodologies may be difficult in peer-reviewed journals, as noted also by \cite{CampanarioConsolationScientistSometimes1993a} and \cite{RichiardiCommonProtocolAgentBased2006a}. Second, computer scientists tend to publish papers in conference outlets. Especially when compared to social scientists, this preference can drive the prevalence of conference papers in the present sample \citep{TichyExperimentalevaluationcomputer1995a, FranceschetRoleConferencePublications2010}.
Retrieved records have been published across a total of 160 venues (either a conference or book series or a journal). The venue with the highest number of records is Lecture Notes in Computer Science\footnote{Complete name: Lecture Notes In Computer Science Including Subseries Lecture Notes In Artificial Intelligence And Lecture Notes In Bioinformatics} with 34 articles, followed by Advances in Intelligent Systems and Computing (20) and ACM International Conference Proceeding Series (17), Ceur Workshop Proceedings (7) and Proceedings of SPIE - The International Society for Optical Engineering (7). The five most represented journals are Procedia Computer Science (6), Computer and Security (5), Interfaces (5), International Journal of Innovative Technology and Exploring Engineering (5), and the Russian Journal of Criminology (5). It is worth to note that the publisher of the International Journal of Innovative Technology and Exploring Engineering, namely ``Blue Eyes Intelligence Engineering \& Sciences Publication'' was listed in the last version of the infamous Beall's list of predatory publishers \citep{StrielkowskiPredatoryjournalsBeall2017a}.
{}

Two considerations emerge from these numbers. First, works on AI+crime are sparsely distributed across a heterogeneous and wide number of venues. This indicates that a proper homogeneous subfield of research has not emerged yet and that scientists have not yet found a proper dedicated venue for research dissemination (or that this venue simply does not exist). Heterogeneity and lack of cohesiveness are also demonstrated by the fact that among the most frequent venues (although they each account for 1.32\% of the total venues) are a journal that is allegedly connected to a predatory publisher and that has been indexed by Scopus in 2018 and a non-Western criminology journal that has been founded in 2016. Second, and connected to this latter point, it is interesting to note that Western criminology journals are marginally present in the list of venues (the only Western criminology and criminal justice journals that are reported are ``Crime science'', ``Journal of Criminal Justice Education'', and the ``Journal of Quantitative Criminology''). This may suggest that specialized journals in these fields may not be ready to embrace sophisticated new methods derived from computer science and AI. Alternatively, it may be that authors working at the intersection between crime and AI are prominently from fields other than criminology and criminal justice, and potentially mainly from computer science, thus making criminology journals less attractive for their careers and research aims.

\subsection{Graphs of Author- and Index- Keywords: Patterns of Themes and Topics}
\begin{figure}[h!]
    \centering
    \includegraphics[scale=0.5]{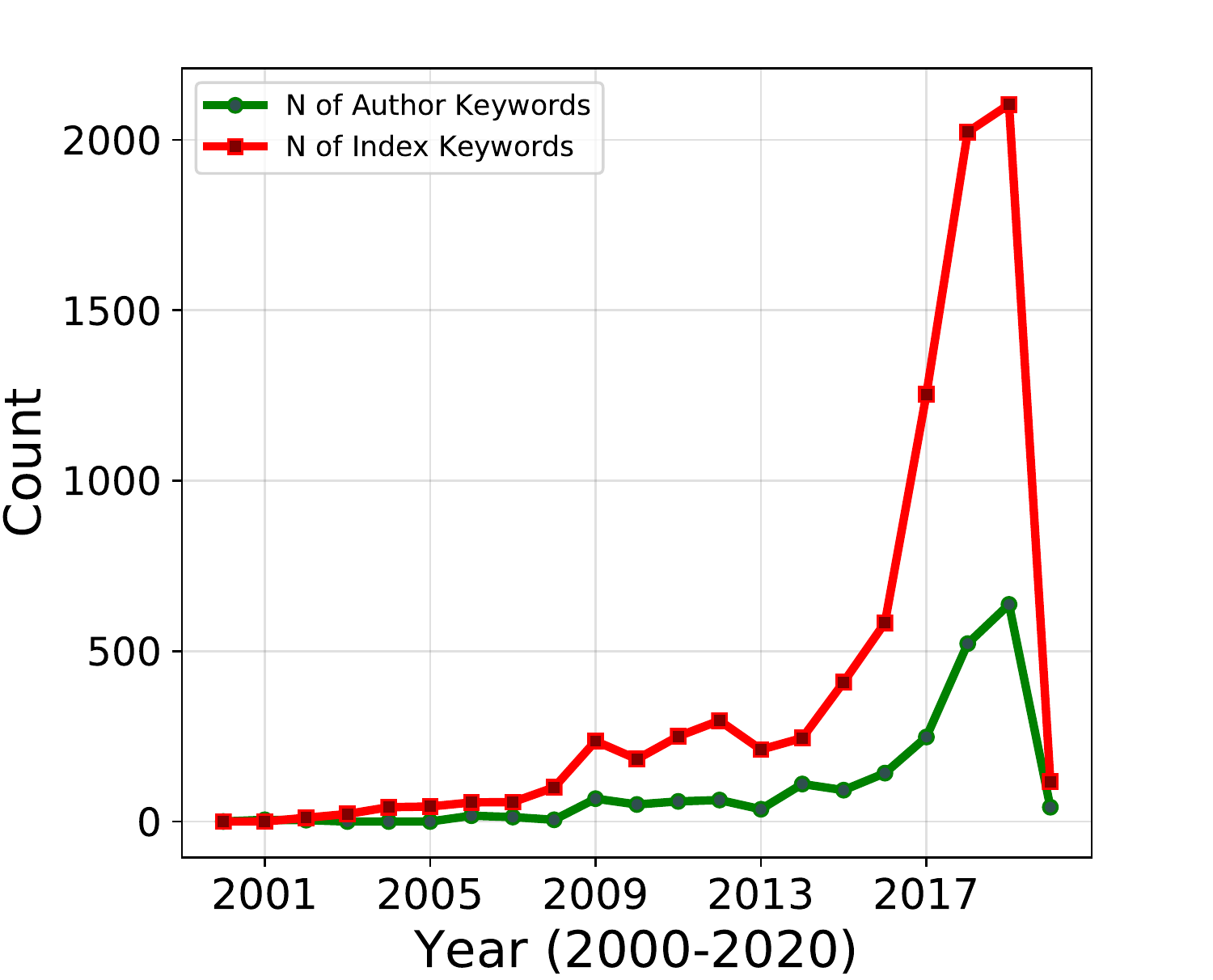}
    \caption{Yearly Trends of Author and Index Keywords}
    \label{fig:keytrends}
\end{figure}

Keywords are a useful variable to measure the evolution of scientific production. This also applies to the literature at the intersection between AI and crime. Figure \ref{fig:keytrends}\footnote{The graph only reports the trends from 2000 on because for items published before 2000 information on author keywords was largely missing.} shows the temporal trends of keywords in the last twenty years. The plot highlights how, as the number of publications increases, so do the number of authors and index keywords. The higher number of index keywords is driven by the fact that Scopus does not bound them to a fixed quantity, while authors usually have a maximum number of keywords to be listed in their publications. Overall, such figures suggest that not only the interest of researchers for AI applications for crime-related problems has sensibly grown over the past two decades. It also indicates that the number of topics, algorithms, and problems being investigating is augmenting over time. The yearly increase in the size of the literature on AI and crime is followed by a parallel growing heterogeneity of research problems.
\newpage

In order to try to understand what are the most common topics investigated in this area, keywords have been processed as to create graphs of co-occurrence across publications. Using all the keywords (both index and author ones) included in the dataset, two separate matrices of co-occurrence was created. Two distinct graphs, in the form $G=(V,E,W)$, where $V$ is the set of nodes (keywords), $E$ is the set of edges mapping connections (co-occurrence across publications) among keywords and $W$ maps the set of weights associated to each edge (namely, the number of times two keywords are related). Table \ref{net_compar} highlights the most important features of the two networks as a whole. 

What immediately emerges from the table is that the two graphs have sensibly different characteristics. The Index keyword graph has many more nodes (i.e., keywords) and edges, also in proportion to the Author-keyword graph, resulting in a higher density of the given network as a whole. In relation to this, the Author-keyword graph has longer characteristic path length and diameter compared to the Index-keyword graph, suggesting that the former is much more sparse and disconnected. The disconnectedness of the graph is testified also by the value of network fragmentation, which map the proportion of nodes that are disconnected in the whole set. As it can be seen, the Author-Keyword graph includes a considerable number of small components (1 isolate, 3 dyads, 19 triads), while the Index-Keyword Graph has only two components: a dyad, and the core one which accounts for 99.999\% of the total of nodes. \\
\newpage
\begin{table}[!hbt]
\centering
\footnotesize
\caption{Comparison of Network Features and Metrics for Graphs of Author Keywords and Index-Keywords}
\begin{tabular}{lcc}
\hline
\textbf{Feature / Metric} & \textbf{\begin{tabular}[c]{@{}c@{}}Author Keyword\\ Graph\end{tabular}} & \textbf{\begin{tabular}[c]{@{}c@{}}Index Keyword\\ Graph\end{tabular}} \\\hline\hline
N of Nodes & 1,719 & 3,897 \\\hline
\begin{tabular}[c]{@{}l@{}}N of Edges (Non Self-Loops)\end{tabular} & 11,912 & 117,340 \\\hline
Density & 0.005 & 0.008 \\\hline
\begin{tabular}[c]{@{}l@{}}Characteristic  Path Length\end{tabular} & 3.652 & 2.497 \\\hline
\begin{tabular}[c]{@{}l@{}}Diameter (Only Reachable Pairs)\end{tabular} & 10 & 6 \\\hline
\begin{tabular}[c]{@{}l@{}}Network Fragmentation\end{tabular} & 0.272 & 0.001 \\\hline
Isolates & 1 & 0 \\\hline
Dyads & 3 & 1 \\\hline
Triads & 19 & 0 \\\hline
\begin{tabular}[c]{@{}l@{}}Larger Components (\textgreater{}4)\end{tabular} & 38 & 1 \\\hline
Mean (St. Dev.) & 43.55 (238.85) & 3,895 (0.00)\\\hline
\end{tabular}
\label{net_compar}
\end{table}

\begin{figure}[!hbt]
\centering
\begin{subfigure}{.47\textwidth}
  \centering
  \includegraphics[scale=0.40]{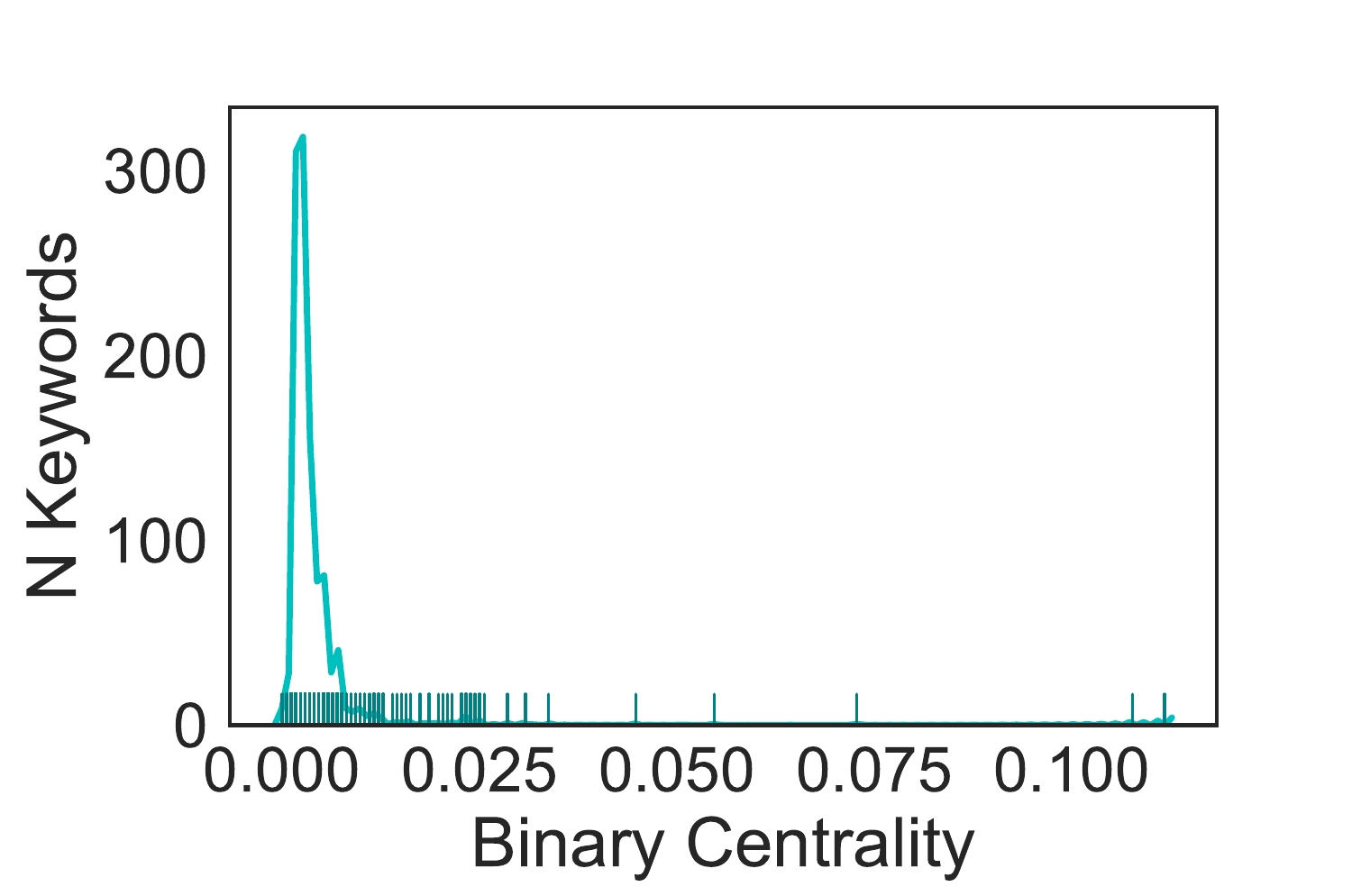}
  \caption{Author Keyword Graph}
  \label{fig:sub1}
\end{subfigure}%
\begin{subfigure}{.47\textwidth}
  \centering
  \includegraphics[scale=0.40]{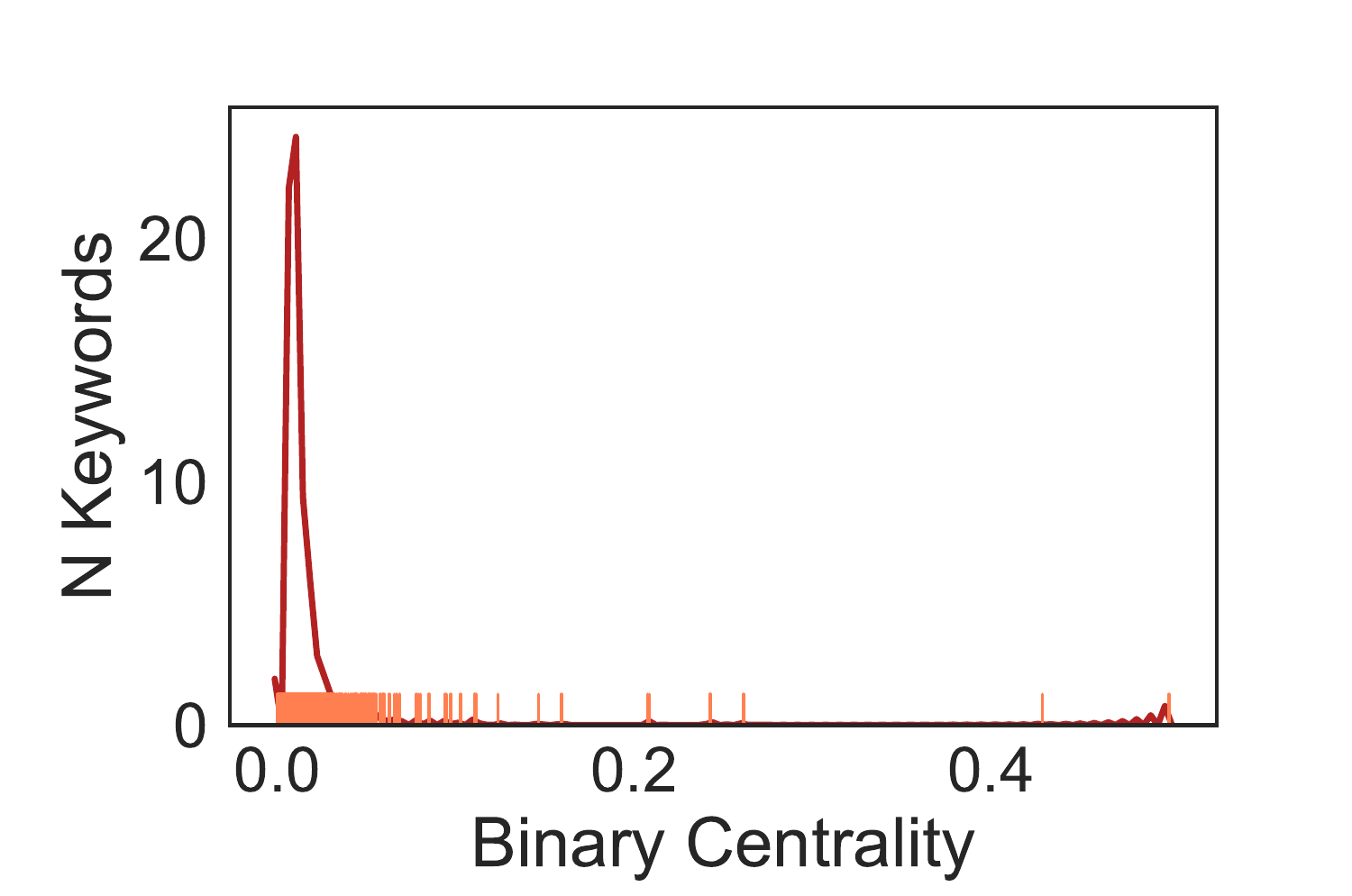}
  \caption{Index Keyword Graph}
  \label{fig:sub2}
\end{subfigure}
\caption{Kernel Density Estimation of Binary Centrality Values Distribution}
\label{comparison}
\end{figure}

These differences in the graphs are due to the distinct nature of the keywords used to characterize each publication. Author keywords are much more discretionary, as the choice is completely left to the authors, while in the case of Index keywords, the procedure is much more standardized and it is carried out by professional indexers based on several available thesauri. On one hand, notwithstanding the higher number of keywords (i.e., nodes in the graph), Index keywords are more densely connected and may be less useful in capturing existing patterns in publications. On the other hand, the standardized procedure employed for categorizing studies by Scopus reduces the issue of having the same words written differently (e.g., with capital letters, in British or American English). Figure \ref{comparison} shows the kernel density estimation and the distribution of the centrality values of in the binarized author keyword and index keyword graphs. Author keywords are much more clustered around values very close to zero, further highlighting the sparseness of topics. When index keywords are considered, the picture sensibly changes, despite a prominent left-skewness of the distribution. Index keywords, compared to author ones, are more densely connected.

Tables \ref{2}, \ref{3}, \ref{4}, \ref{5} focus on the most central keywords in both the author and index keyword graphs. Tables \ref{2} and \ref{3} specifically consider the ten most central keywords overall. Table \ref{2} demonstrates the very high popularity of \textit{Machine Learning} as a keyword used by authors in their work, with a centrality of 0.26 (meaning that 26.7\% of the whole set of 1719 keywords chosen by authors are associated with \textit{Machine Learning}). The broader expression \textit{Artificial Intelligence} is the second-most central word, followed by \textit{Deep Learning} and \textit{Data Mining}. \textit{Classification} appears to be the most common performed task by scholars in the sample, as it is ranked fifth in the overall list, and \textit{Random Forest} and \textit{Neural Networks} are the two most popular classes of algorithms (and the only ones present within this specific list).\footnote{\color{black} When removing the three most central keywords in the graph, which are overlapping with the keywords included in the query, the ranking remains the same, with Data Mining \color{black} becoming the most central (0.072), followed by Big Data (0.043) and Crime Prediction (0.032).} \\
The most central keywords in the index graph are partially overlapping with the ones found in the author ranking. \textit{Learning Systems} is the most popular (50.2\% of the keywords are associated with this particular keyword), followed by \textit{Crime}. \textit{Artificial Intelligence}, \textit{Machine Learning} and \textit{Deep Learning}, the three AI-related expressions used for the search query, are ranked third, fifth, and seventh respectively. \textit{Classification (of Information)} is ranked tenth and further indicates the prevalence of classification tasks within the sample of works retrieved from Scopus.

While tables \ref{2} and \ref{3} reported the most central keywords overall, tables \ref{4} and \ref{5} specifically report the ten most central crime-related keywords. Crime-related keywords are keywords that are connected somehow with criminal phenomena, criminology areas of criminological topics. From both tables emerges the prevalent interest of scientists for cyber-related topics. In Table \ref{4}, \textit{Cybercrime} is ranked third, \textit{Malware} is fifth, \textit{Security}, a word which is generally connected with the cyber-sphere, is sixth, \textit{Cyber Security} and \textit{Phishing} are ranked seventh and eighth respectively. Finally, although \textit{Fraud Detection} is not inherently cyber-related, many applications in fraud detection studies encompass digital or computer-related frauds. In spite of different specific keywords, the picture is substantially similar in the index graph. \textit{Computer crime} is the second-most central keyword, followed by \textit{Network Security} and \textit{Malware}. Other cyber-related popular keywords are \textit{Security Systems} and \textit{Intrusion Detection}. 

These results provide a clear picture of the most trending topics in the area at the intersection between AI and crime. What I have broadly defined as cyber-related topics are extremely popular across both graphs, and their prevalence is even more evident considering the almost complete absence of keywords related to other areas of crime and criminology (with the exception of \textit{Criminal Law} in the author graph and \textit{Law Enforcement} and \textit{Forensic Science} in the index graph).

\begin{table}[!hbt]
\setlength{\tabcolsep}{2pt} 
\renewcommand{\arraystretch}{0.9}
    \footnotesize
    \begin{minipage}{.5\textwidth}
    \caption{Most Central Keywords -\\ Author (Overall)}
      \centering
      \begin{tabular}{clc}
            \hline
            \multicolumn{1}{l}{\textbf{Rank}} & \textbf{Keyword} & \textbf{\begin{tabular}[c]{@{}c@{}}Binary \\ Centr.\end{tabular}} \\\hline\hline
            1 & Machine Learning & 0.267 \\
            2 & Artificial Intelligence & 0.111 \\
            3 & Deep Learning & 0.107 \\
            4 & Data Mining & 0.073 \\
            5 & Classification & 0.054 \\
            6 & Big Data & 0.044 \\
            7 & Crime Prediction & 0.033 \\
            8 & Random Forest & 0.030 \\
            9 & Neural Networks & 0.030 \\
            10 & Crime & 0.028\\\hline
    \end{tabular}

    \label{2}
    \end{minipage}
    \begin{minipage}{.5\textwidth}
    \setlength{\tabcolsep}{2pt} 
    \renewcommand{\arraystretch}{0.9}
      \centering
      \caption{Most Central Keywords - \\ Index (Overall)}
      \begin{tabular}{clc}
      \hline
            \textbf{Rank} & \textbf{Keyword} & \textbf{\begin{tabular}[c]{@{}c@{}}Binary\\ Centr.\end{tabular}} \\\hline\hline
            1 & Learning Systems & 0.502 \\
            2 & Crime & 0.499 \\
            3 & Artificial Intelligence & 0.428 \\
            4 & Learning Algorithms & 0.261 \\
            5 & Machine Learning & 0.242 \\
            6 & Computer Crime & 0.242 \\
            7 & Deep Learning & 0.208 \\
            8 & Data Mining & 0.207 \\
            9 & Neural Networks & 0.159 \\
            10 & Classification (of Information) & 0.146\\\hline
    \end{tabular}
    \label{3}
    \end{minipage}
  \end{table}

\begin{table}[!hbt]
\setlength{\tabcolsep}{2pt} 
\renewcommand{\arraystretch}{0.9}
    \footnotesize
    \begin{minipage}{.5\textwidth}
    \caption{Most Central Keywords - \\ Author (Crime-related)}
      \centering
\begin{tabular}{clc}
\hline
\textbf{\begin{tabular}[c]{@{}c@{}}Rank \\ (Overall)\end{tabular}} & \textbf{Keyword} & \textbf{\begin{tabular}[c]{@{}c@{}}Binary \\ Centr.\end{tabular}} \\\hline\hline
1 (7) & Crime Prediction & 0.033 \\
2 (10) & Crime & 0.028 \\
3 (16) & Cybercrime & 0.023 \\
4 (18) & Crime Analysis & 0.023 \\
5 (21) & Malware & 0.022 \\
6 (24) & Security & 0.022 \\
7 (25) & Cyber Security & 0.021 \\
8 (26) & Phishing & 0.020 \\
9 (32) & Fraud Detection & 0.016 \\
10 (36) & Criminal Law & 0.015\\\hline
\end{tabular}

    \label{4}
    \end{minipage}
    \begin{minipage}{.5\textwidth}
    \setlength{\tabcolsep}{1.7pt} 
    \renewcommand{\arraystretch}{0.9}
      \centering
     \caption{Most Central Keywords - \\
    Index (Crime-related)}
\begin{tabular}{clc}
\hline
\textbf{\begin{tabular}[c]{@{}c@{}}Rank \\ (Overall)\end{tabular}} & \textbf{Keyword} & \textbf{\begin{tabular}[c]{@{}c@{}}Binary \\ Centr.\end{tabular}} \\\hline\hline
1 (2) & Crime & 0.499 \\
2 (6) & Computer Crime & 0.242 \\
3 (15) & Network Security & 0.110 \\
4 (21) & Malware & 0.084 \\
5 (22) & Security Systems & 0.080 \\
6 (25) & Law Enforcement & 0.068 \\
7 (26) & Criminal Investigation & 0.068 \\
8 (30) & Intrusion Detection & 0.062 \\
9 (32) & Criminal Activities & 0.059 \\
10 (34) & Forensic Science & 0.059\\\hline
\end{tabular}
    \label{5}
    \end{minipage}
  \end{table}

Two complementary explanations could help in decoding the central role of cyber-related keywords in both graphs. First, cyber-related topics, which are fairly recent compared to other criminal phenomena, have witnessed a constantly growing interest of researchers enhanced by the inherent hybrid nature of crimes belonging to this sphere (both humans and machines are involved), naturally favored trans-disciplinary research across domains such as criminology and computer science. Second, datasets for cyber-related crimes are generally much wider and richer compared to other data sets recording information for other crimes (e.g., robberies), potentially due to the intrinsic digital nature of crimes occurring in the cyber domain. This would, therefore, facilitate data availability for scientists.

In spite of the vibrant debate around algorithmic decision-making processes in policing and criminal justice, this analysis highlights the peripheral role of keywords associated to these two areas in the sample.\footnote{\textit{Predictive Policing} is ranked 43rd in the author graph and 505th in the index graph. \textit{Criminal Justice System} is ranked 400th in the author graph and 318th in the index graph.} 
Similarly, keywords related to extremely critical and relevant topics such as transparency, bias, fairness, and ethics are also peripheral in both graphs, suggesting that, so far, researchers are more interested in applications rather than societal and ethical implications of research that applies AI algorithms to crime-related issues. \textit{Transparency} is ranked 71st in the author graph and 398th in the index one. \textit{Bias} is ranked 918th in the author graph and \textit{(Intrinsic) Bias} 1481st in the index one. \textit{Fairness} is ranked 151st in the author graph and 1370th in the index one. Finally, \textit{Ethics} is ranked 152nd in the author graph and \textit{(Codes of) Ethics} is 3674th in the index one.\footnote{It is worth to note that the highest-ranked keyword related to the four topics cited in the text are here reported. \textit{Bias}, for instance, can be related to \textit{Machine Bias}, \textit{Gender Bias}, \textit{Algorithmic Bias}, etc.: only the keyword which has the higher centrality is reported for the sake of brevity, meaning that non-listed ones are even more peripheral.} \color{black} Interestingly, the very few papers addressing these problems in the sample have been mostly published in 2018 and 2019, showing that the research community has only very recently started to reason about these issues. \color{black}

\begin{figure}[!hbt]
\centering
\begin{subfigure}{.5\textwidth}
  \centering
  \includegraphics[scale=0.45]{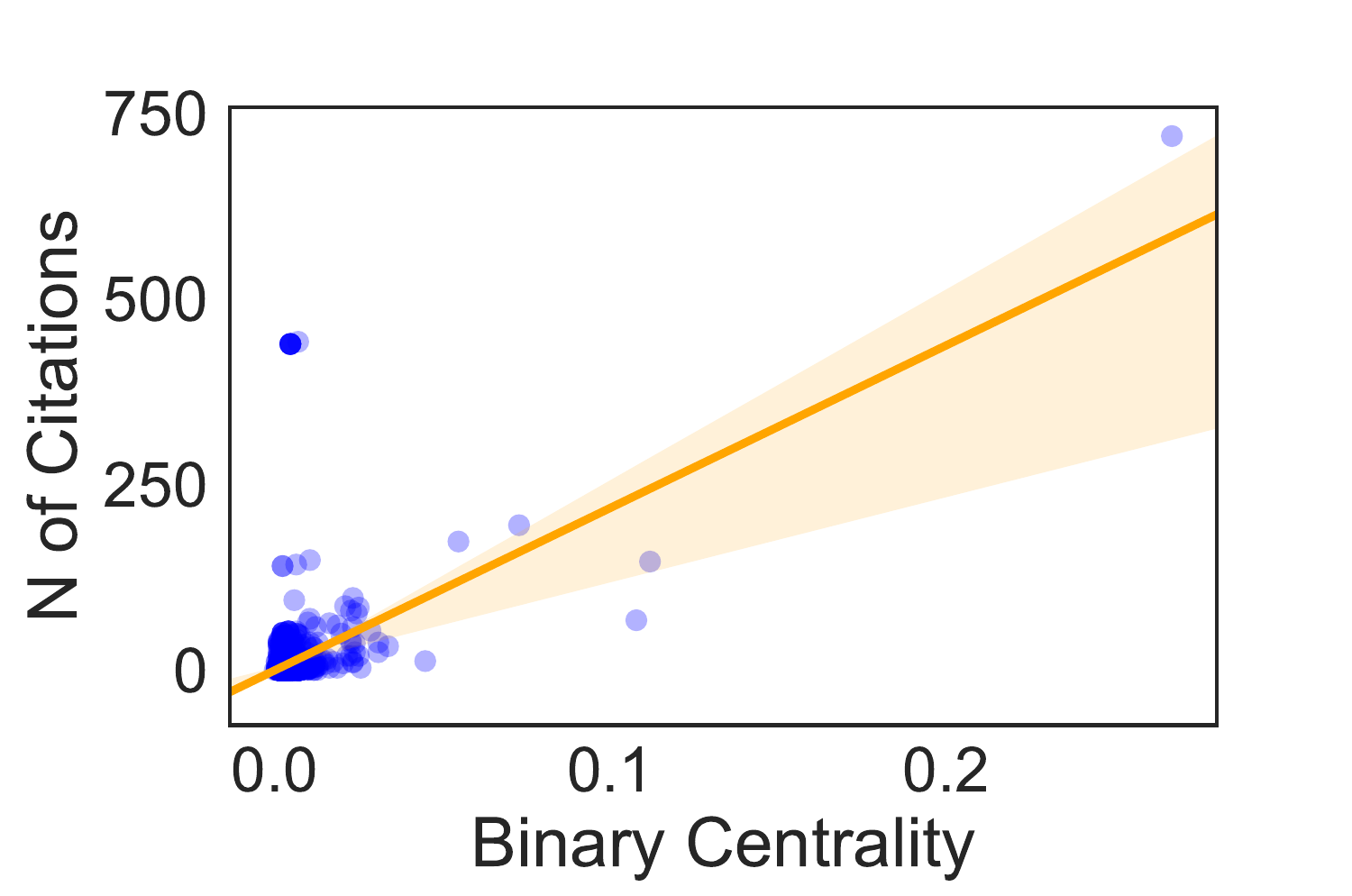}
  \caption{Author Keyword Graph \\
  Pearson's $r$=0.45\\$p$-val $<$ 0.0001}
  \label{fig:sub1cen}
\end{subfigure}%
\begin{subfigure}{.5\textwidth}
  \centering
  \includegraphics[scale=0.43]{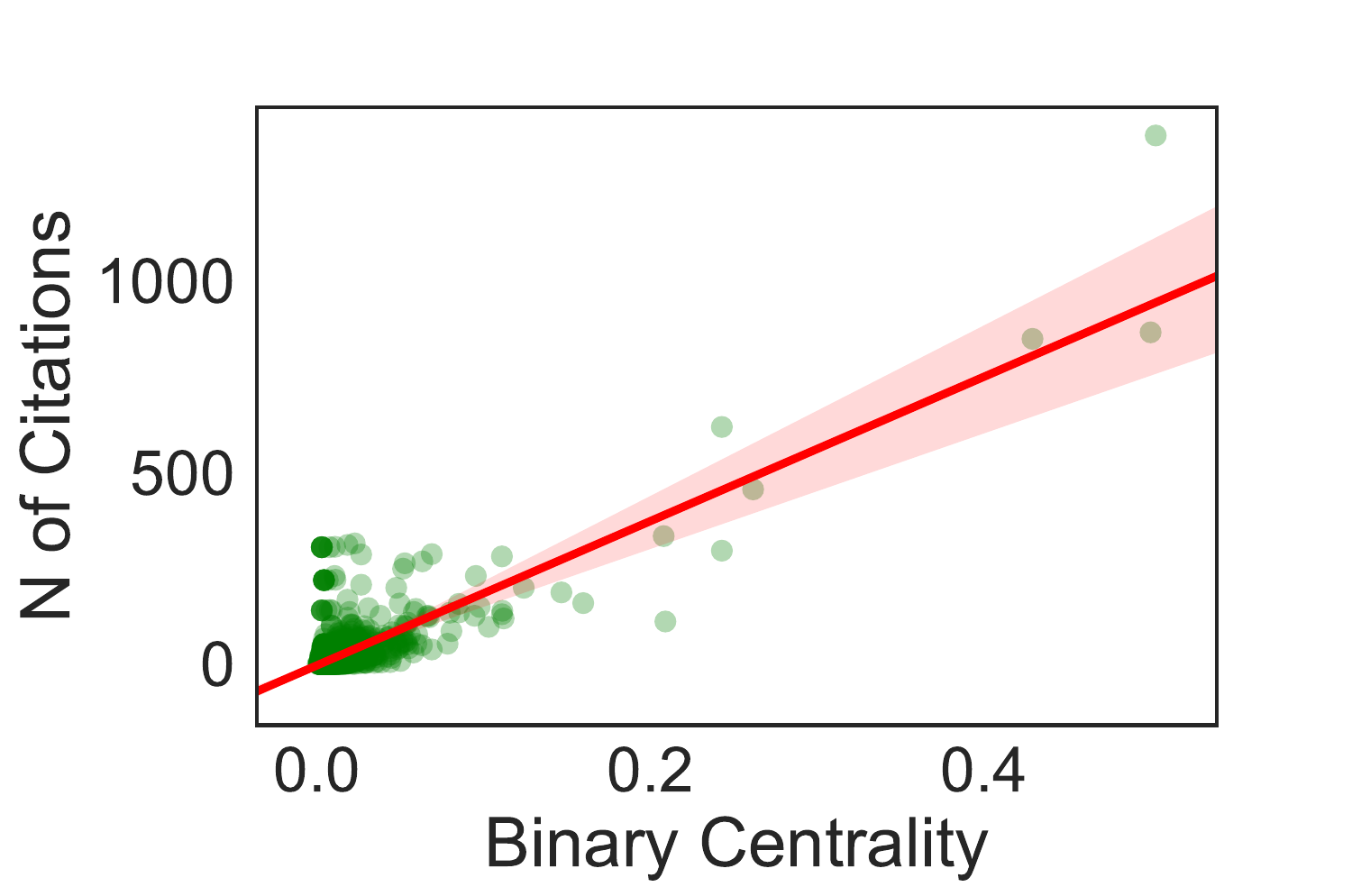}
  \caption{Index Keyword Graph\\
  Pearson's $r$=0.78\\$p$-val $<$ 0.0001}
  \label{fig:sub2cen}
\end{subfigure}
\caption{Bivariate Relation Between Binary Centrality and Number of Citations}
\label{bivariate_comp}
\end{figure}

The network-based analysis of keywords co-occurrence is relevant to detect the most common areas upon which researchers are focusing but can also be interesting in highlighting what could be likely developments in the future. In fact, as Figures \ref{fig:sub1cen} and \ref{fig:sub2cen} show, there is a clear relationship between the centrality of a certain keyword and the sum of times works using that given keyword have been cited. Furthermore, after calculating the prevalence of each keyword (namely the share of the number of papers in which a given keyword is used out of the total of works in the sample), data reveal an almost overlapping positive relation also between citation count and prevalence. 
This interestingly relates to the finding commented above regarding specific themes or topics that are not yet particularly popular in works at the intersection between AI and crime, especially when compared to the whole universe of keywords employed either by authors or professional indexers.
\begin{figure}[!htb]
\centering
\begin{subfigure}{.5\textwidth}
  \centering
  \includegraphics[scale=0.43]{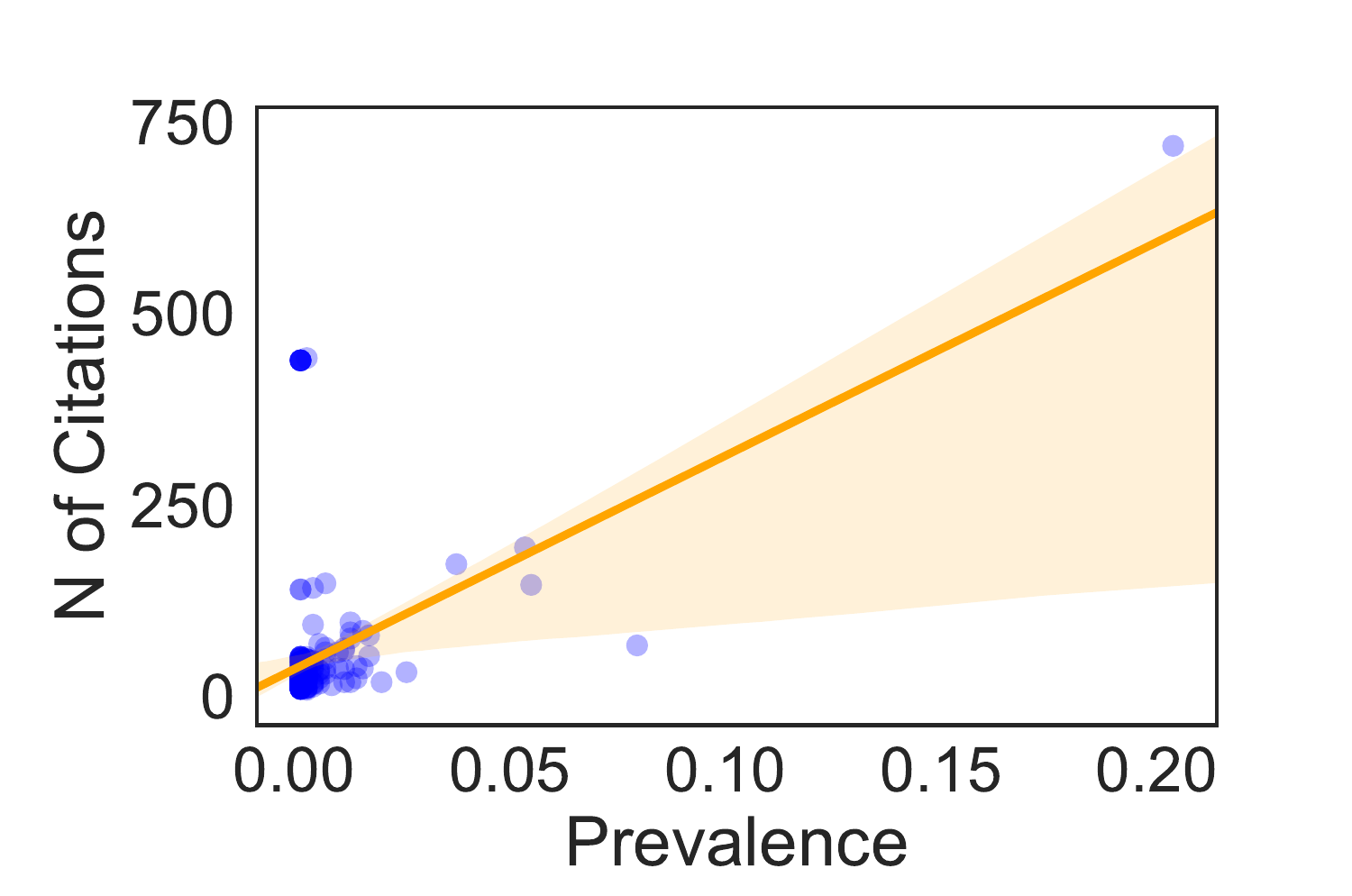}
  \caption{Author Keyword Graph \\
  Pearson's $r$=0.45. \\$p$-val $<$ 0.0001}
  \label{fig:sub1prev}
\end{subfigure}%
\begin{subfigure}{.5\textwidth}
  \centering
  \includegraphics[scale=0.45]{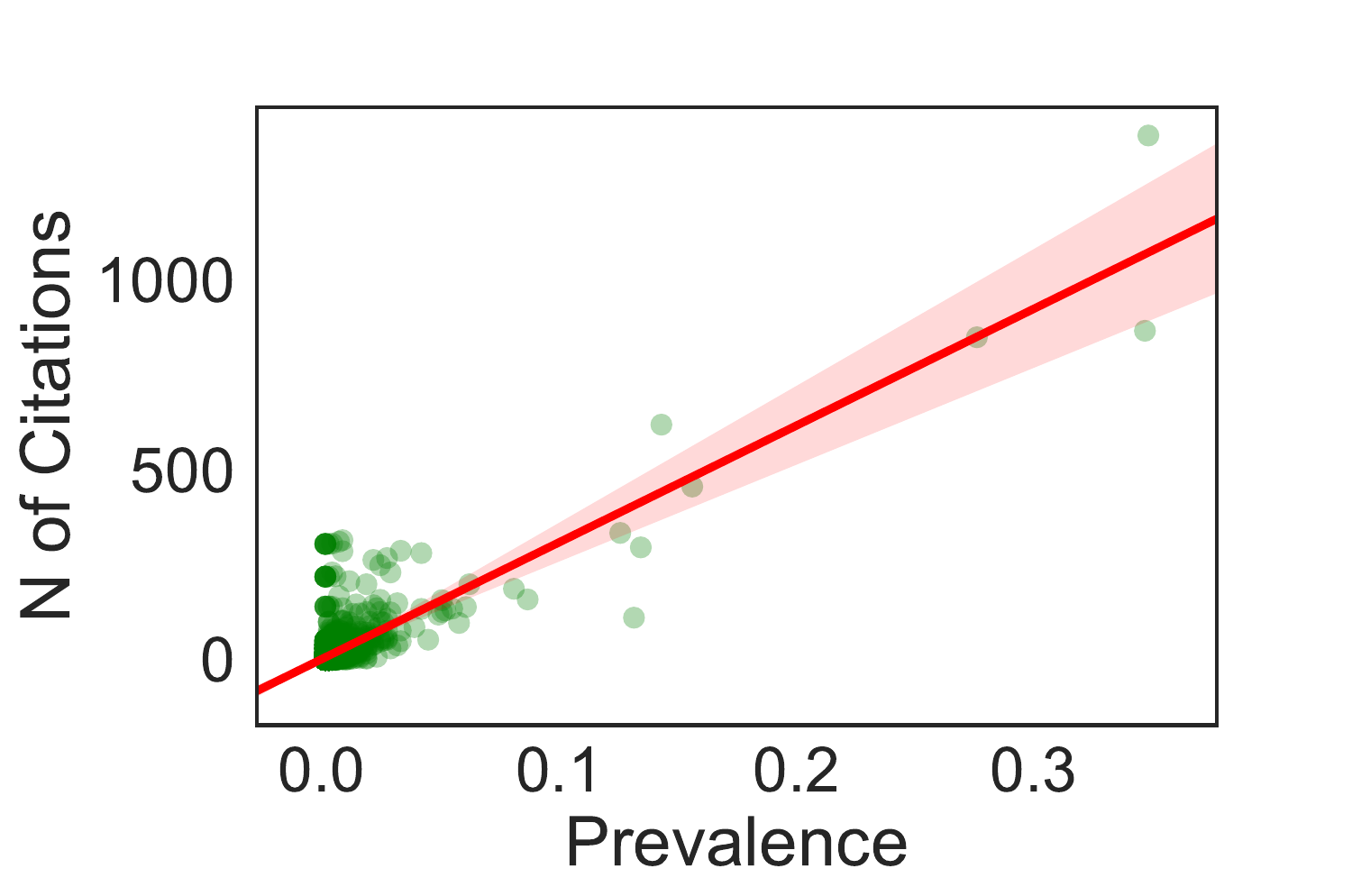}
  \caption{Index Keyword Graph\\
  Pearson's $r$=0.79. \\$p$-val $<$ 0.0001}
  \label{fig:sub2prev}
\end{subfigure}
\caption{Bivariate Relation Between Keyword Prevalence and Number of Citations}
\label{bivariate_comp_prev}
\end{figure}
\newpage

\subsection{Graphs of Collaboration: Authors and Countries}

\begin{figure}[!htb]
    \centering
    \includegraphics[scale=0.65]{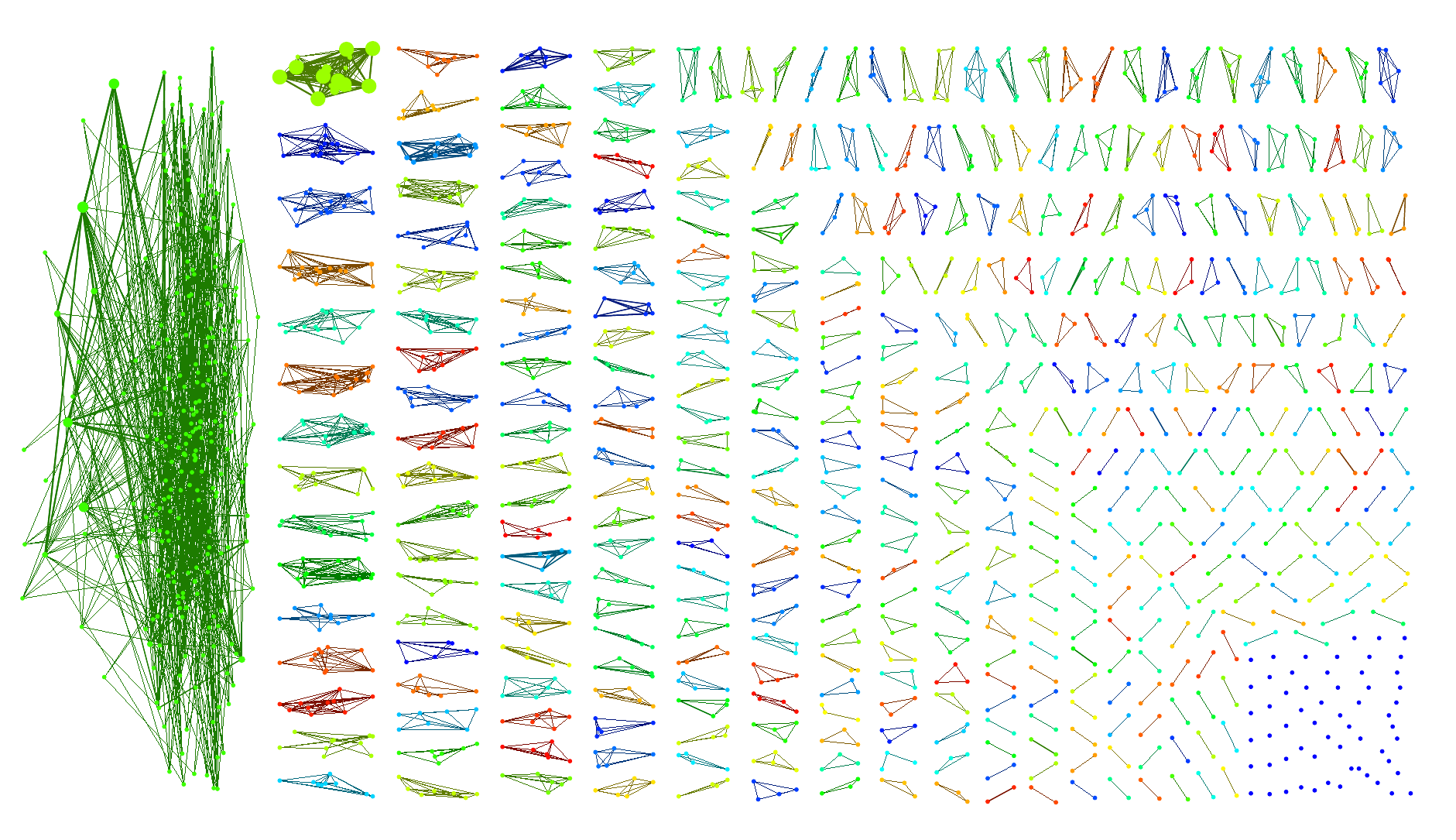}
    \caption{Graph of Authorship and Scientific Collaboration - Component Layout. Links are log-scaled in terms of width (number of collaborations) and Nodes are log-scaled in terms of total degree centrality}
    \label{fig:net_aut_plot}
\end{figure}

The degree to which authors are connected through publication co-authorship is another relevant way to map the state of a specific academic area. With this regard, analyzing the graph of co-authorship of scholars that have authored publications in the present sample can enhance our understanding of scientific inquiry at the intersection between crime and AI. Figure \ref{fig:net_aut_plot} illustrates the co-authorship network with a component layout that highlights the different groups of researchers collaborating together. What immediately emerges is that the graph is particularly disconnected, with one large component consisting of only 227 authors accounting for 13.59\% of the total number of scholars in the sample. 

\color{black} Compared to other studies and taking into consideration the dimension of the larger component, the disconnectedness of the graph is straightforward. In his seminal studies, Newman \cite{Newmanstructurescientificcollaboration2001, NewmanCoauthorshipnetworkspatterns2004} suggested the importance of giant components in facilitating the flow of ideas through faster communication and easier access to collaboration. Contrarily, the absence of a giant component indicates an immature and poorly cohesive research community. Concerning computer science, for instance, Elmaciouglu and Lee \cite{Elmacioglusixdegreesseparation2005} investigated the co-authorship structure using data from the DBLP computer science bibliography from 1968 to 2013 and calculated that the giant component in the network accounted for more than 57\% of the authors. Using information from the same archive, Franceschet \cite{FranceschetCollaborationcomputerscience2011} as well indicated the presence of giant component as a sign of cohesiveness, coupled with high assortativity.
Similarly, Huang and Li \cite{HuangCollaborationTimeCharacterizing2008} found a giant component in the co-authorship network retrieved from the CiteSeer database (from 1988 to 2005). While a study by Moody \cite{MoodyStructureSocialScience2016} determined the existence of a giant component in sociology as well, a more recent work by Gonzalez-Alcaide et al. \cite{Gonzalez-AlcaideProductivityCollaborationScientific2013} highlighted instead how, in a sample of criminology articles, 78.5\% of the co-authorship groups found consisted of only two or three authors. This indicated, rather than immaturity, a community in which, according to the authors ``the predominance of work is for a reduced nuclei of researchers'' \cite[p.31]{Gonzalez-AlcaideProductivityCollaborationScientific2013}. In the case of the present work, however, the motivation behind the degree of disconnectedness of the network of contributors at the intersection of AI and crime is rather caused by the fact that the field has started to develop only recently.

\begin{figure}
    \centering
    \includegraphics[scale=0.65]{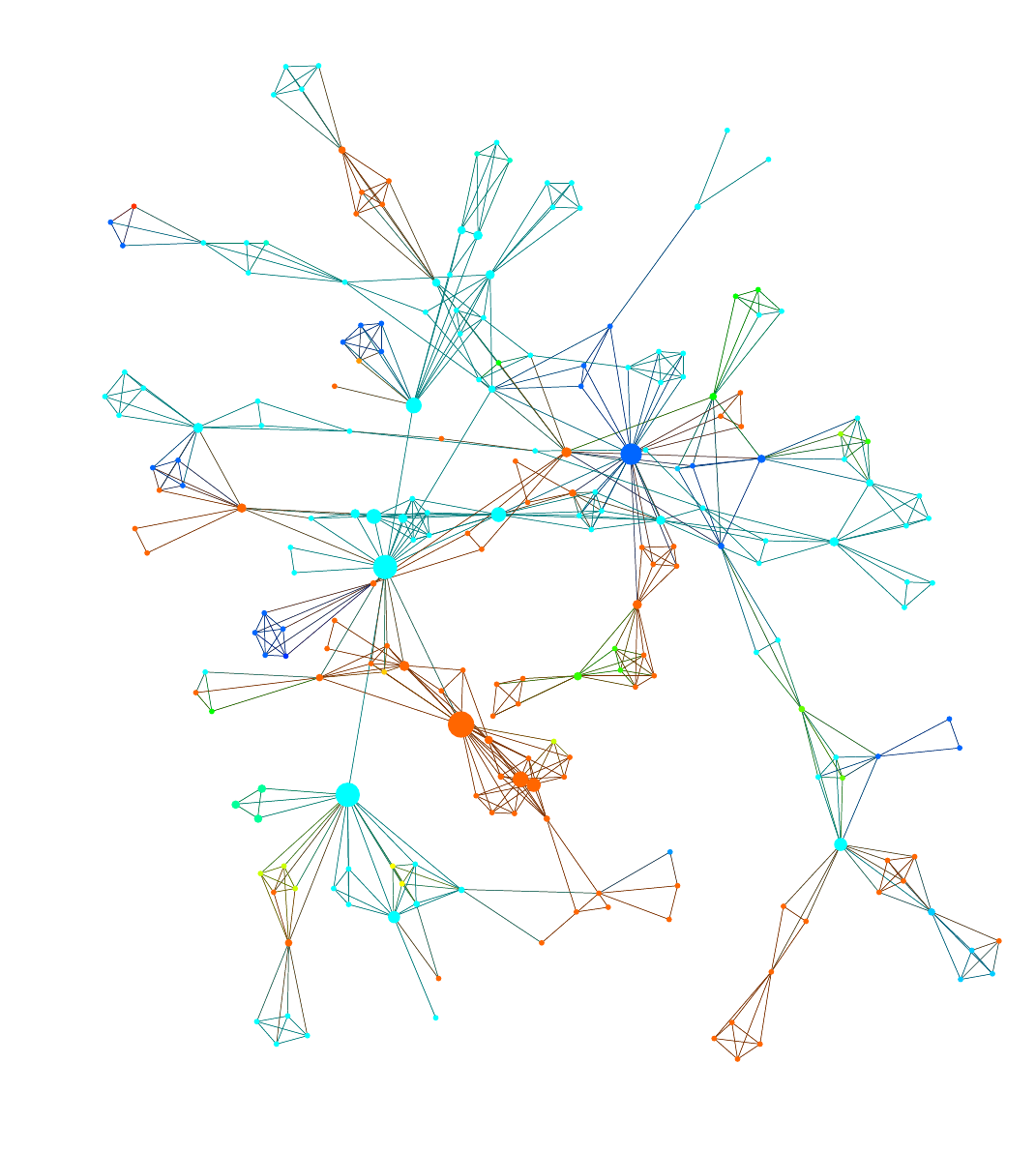}
    \caption{ \color{red}Visualization of the Large Component of the Authorship and Scientific Collaboration Graph. Nodes are scaled in terms of total degree centrality and colored based on the country of affiliation of the author. Orange represents the United States, blue is Australia, light blue represent China-affiliated scholars. Other countries represented in the component are Bangladesh, Canada, Estonia, Hong Kong, India, Macau, Malaysia, Netherlands, South Korea, Spain, and United Kingdom.}
    \label{fig:big component}
\end{figure}

A further investigation of the structure of the larger components (Figure \ref{fig:big component}) shows how the wide majority of scholars within the sub-network are affiliated to institutions based in the United States, China, and Australia. Other countries also appear in the component but marginally in terms of centrality and frequency. The general disconnectedness of the network and the homogeneity of scholars belonging to the larger component in the co-authorship graphs certainly point in the direction of a fragmented area of research, in which scientific production tends to be clustered in few countries. 
\color{black}

Table \ref{aut_net} quantitatively pictures the structure of the overall co-authorship graph. The sample includes a total of 1,964 authors. The density of the network is extremely low, with a total of 59 authors that are isolates. The same pattern is found when focusing on the number of dyads and triads. Overall, 134 dyads and 115 triads are present in the affiliation graph, meaning that more than 30\% of researchers are connected either to a single or two other authors. The larger component includes 227 different researchers.

The very high degree of disconnectedness of researchers working in this area represents a relevant finding. The sparseness of individual-collaboration and the tendency to work in siloed-groups describes a situation in which the circulation of new ideas and the inclusiveness of research projects are not favored. Furthermore, this structure is likely connected to the presence of ``transient'' researchers that contribute to a particular research area only by publishing one or very few papers \citep{GordonTransientcontinuantauthors2007}. Whatever the causal relation (if any) between disconnectedness and the presence of transient researchers, the coupling of these two phenomena discourages the formation of new theories, the replication of research findings, and the development of a homogeneous corpus of literature. 
\newpage
\begin{table}[!htb]
\footnotesize
\centering
\caption{Graph of Co-Authorship - Data and Metrics}
\begin{tabular}{lc}
\hline
\textbf{Graph Feature} & \textbf{Value} \\\hline\hline
N of Nodes (Authors) & 1,964 \\\hline
Density & 0.0028 \\\hline
All edges (Non Self-Loops) & 6,974 \\\hline
N of Isolates & 59 \\\hline
N of Dyads & 134 \\\hline
N of Triads & 115 \\\hline
Larger Components ($>$4) & 191 \\\hline
Max Component & 227 \\\hline
Mean (St.Dev) & 6.76 (16.15)\\\hline
\end{tabular}
\label{aut_net}
\end{table}

\begin{figure}[!htb]
    \centering
    \includegraphics[scale=0.65]{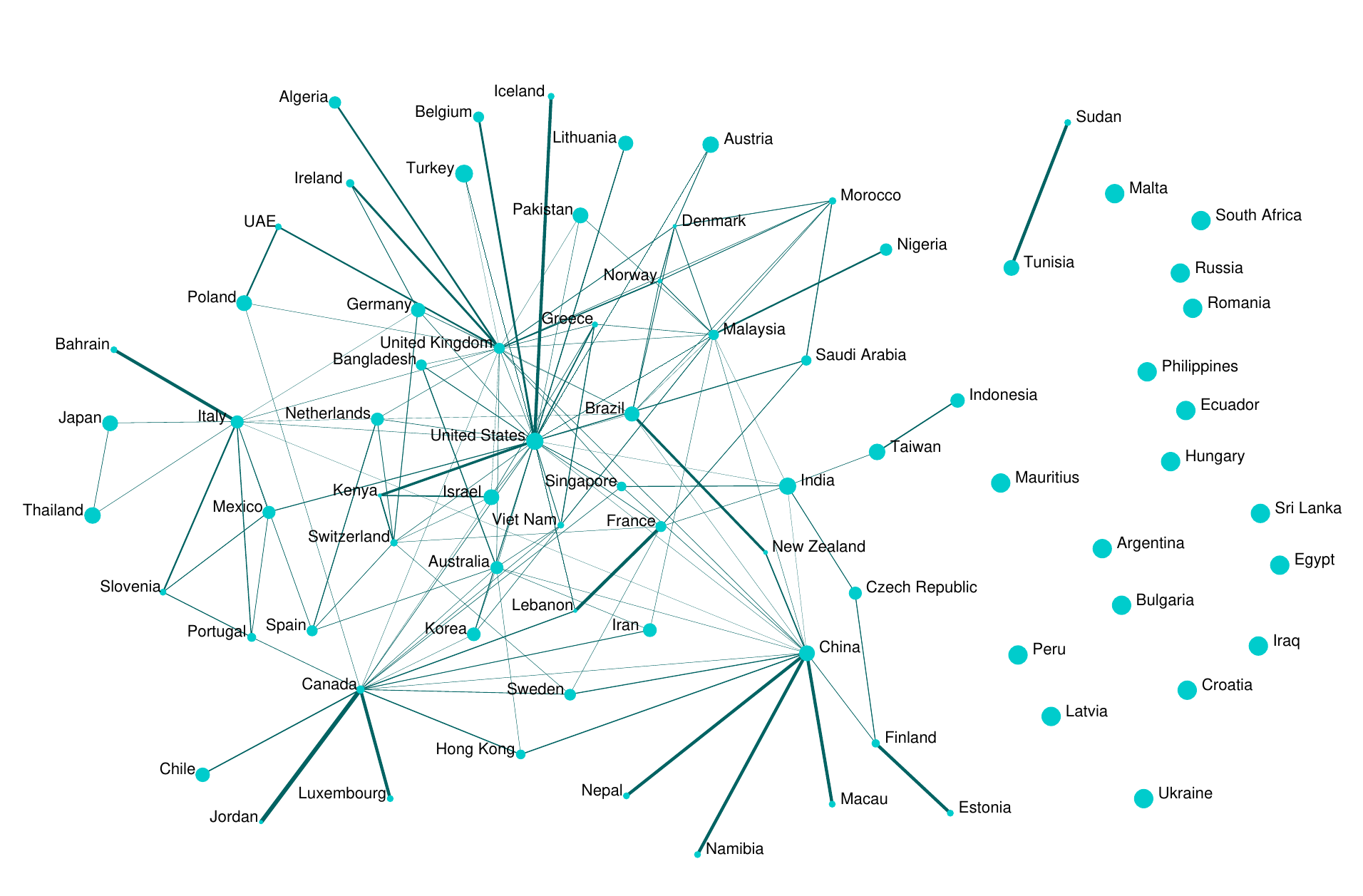}
    \caption{Graph of Scientific Collaboration Among Countries - based on Affiliations. Self-loop (collaborations among authors in the same countries) are excluded. Links are log-scaled in terms of width (number of collaborations). Nodes are log-scaled in terms of the share of domestic collaborations (1 - (share of international collaborations). The larger the node, the lower the number of international collaborations out of the total.}
    \label{fig:countries}
\end{figure}

Further information can be gathered through the analysis of the graph mapping the collaborative relations between countries. Based on the country in which each affiliation (e.g., research lab, group) is based, I have drawn a network where each link quantifies the number of papers published between countries $i$ and $j$. If, for example, researchers $a$ and $b$, based in two different labs located in two countries $i$ and $j$, have published 3 papers together, a weighted link (with weight equal to 3) is created. Figure \ref{fig:countries} displays the graph of relations between countries. Overall, affiliations from 77 countries are present in the entire sample. A total of 17 countries appear as isolated, meaning that no collaboration with foreign affiliation is present in the data.\footnote{These countries are Malta, South Africa, Russia, Romania, Philippines, Ecuador, Hungary, Sri Lanka, Mauritius, Argentina, Bulgaria, Egypt, Iraq, Peru, Latvia, Ukraine, and Croatia.} To inspect the extent to which countries collaborate, I have also calculated the percentage of works with international collaborations out of the total of collaborations.

\begin{figure}[!hbt]
    \centering
    \includegraphics[scale=0.4]{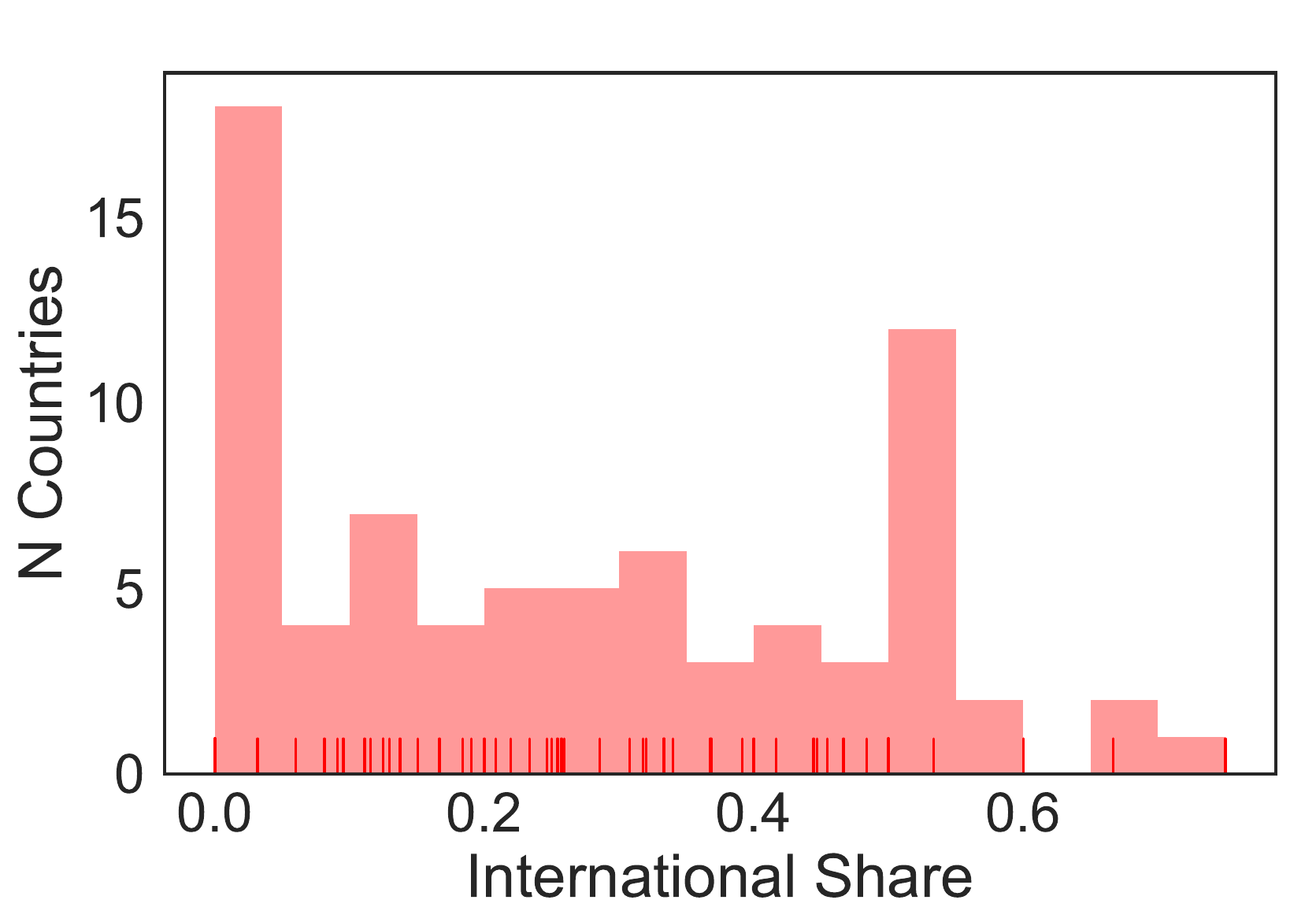}
    \caption{Distribution of International Share of Collaboration in the Sample}
    \label{fig:distint}
\end{figure}

Figure \ref{fig:distint} shows instead the distribution of the international share of collaborations across countries. What emerges is that, on average, there is a low level of international collaboration in terms of publications at the intersection between AI and crime. Besides isolates which trivially only have domestic collaborations, 39 countries (50.64\% of the total) have an international share equal or lower than 0.25, meaning that collaborations are, in at least 3 scientific works out of 4, only between research groups and labs based in that same country. From the international standpoint, the most international countries are Kenya (0.8), Lebanon (0.75), Denmark (0.66),\footnote{The international attitude of Denmark research was also testified in \citep{GlanzelDomesticityinternationalitycoauthorship2005}} Jordan (0.66), Norway (0.6) and New Zealand (0.6). 

\begin{figure}[hbt!]
    \centering
    \includegraphics[scale=0.4]{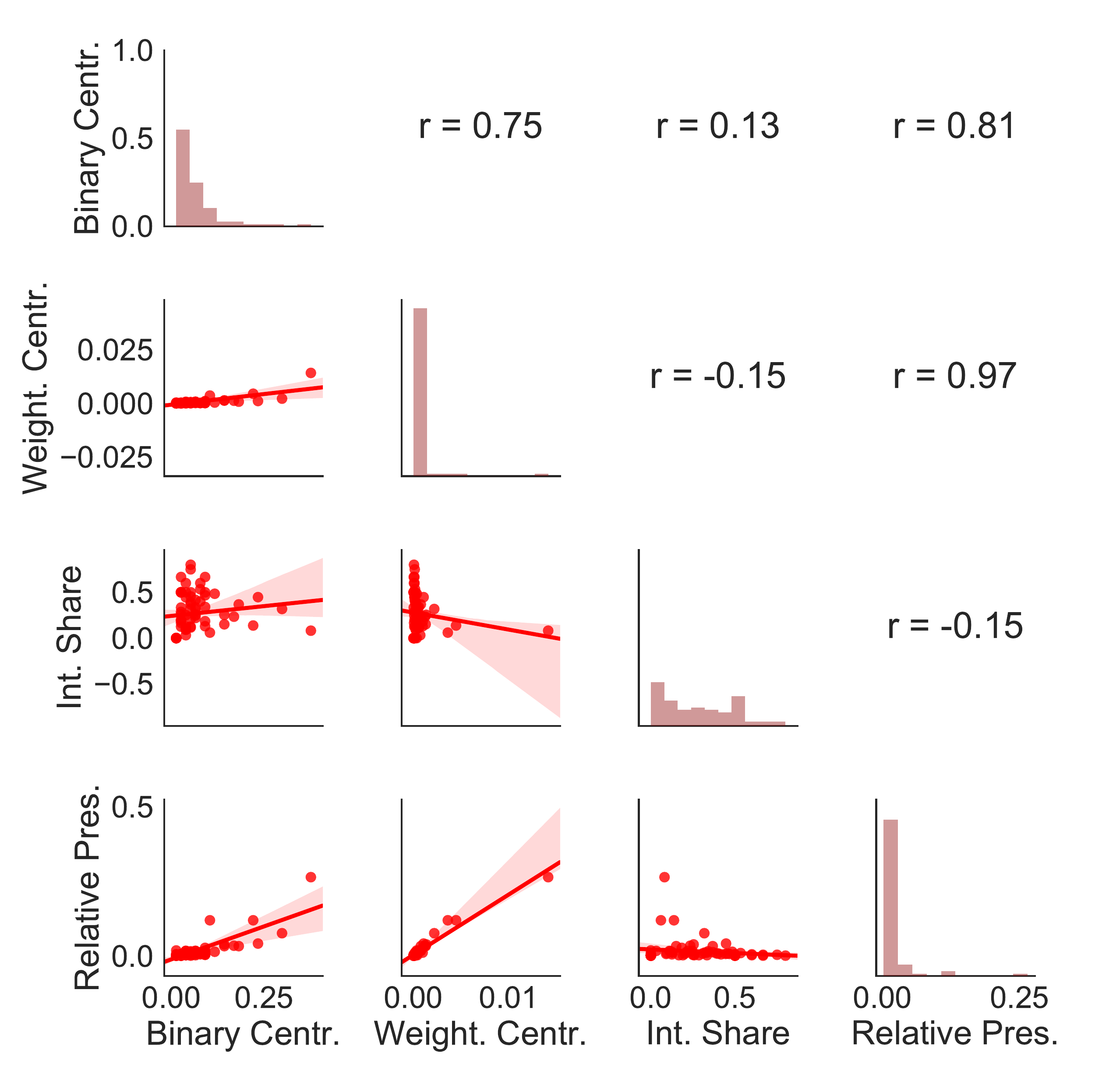}
    \caption{Regression, Distribution and Pearson's Correlation of Binary Centrality, Weighted Centrality, International Share, and Relative Presence for Countries in the Sample. All coefficients are statistically significant at the 99\% confidence level.}
    \label{fig:jointplot}
\end{figure}

To better assess the structure of the graph of international collaborations that emerged from the data, I have also calculated the binary centrality, the weighted centrality, and an indicator of the relative presence of a given country in the sample of 692 works. The binary centrality is simply the normalized centrality in the range [0,1] calculated from the network of binary interactions (i.e., collaborations) between countries. The network is the binarized form of the originally weighted one, in which if two countries $i$ and $j$ have a number of collaborations which is $\geq$ 1, then the entry in the matrix becomes 1, with 0 otherwise.
The weighted centrality is, instead, the normalized centrality computed from the original weighted matrix of collaborations. Finally, the indicator of relative presence simply captures the extent to which a country is represented in the total sample. For a country $i$, the indicator is given by the ratio between the number of works in which at least one author has an affiliation in the country $i$ and the total number of studies, i.e. 692. Some relations emerge (Figure \ref{fig:jointplot}). 

First of all, the relationship between international share and centrality is very low in the binary case and even negative when the weighted matrix is considered. This means that, on average, most central countries tend not to collaborate internationally in this area and that, conversely, collaborations between countries that are peripheral to the network are instead more common. This finding is somehow confirmed also by the negative relation ($r$=-0.15) between international share and relative presence and by the very high correlation ($r$=0.97) between weighted centrality and relative presence. Countries that are very present in the sample\footnote{United States (26.44\%), China (11.99\%), and India (11.99\%) alone are present in more than half of the studies. Other particularly present countries are the United Kingdom (7.65\%), Canada (4.19\%), Australia (3.90\%), and Italy (3.46\%). \color{black} Their prominent role was also highlighted in a work by Hu and Zhang \citep{HuStructurepatternscrossnational2017} focused on the field of Big Data research, which is adjacent to the research area investigated in this paper.} tend to collaborate mostly domestically. This might be related to the unequal distribution of resources and awarded grants across the world, which pushes peripheral countries to create connections with each other to overcome structural inequalities of science. An additional hypothesis may concern the interest of central countries to potentially maintain their knowledge within their borders, especially given the certainly critical blurred area intersecting crime and AI.

Following, Table \ref{continent} lists the average and the standard deviation of the international share values divided per continent. America, which includes 8 countries including the United States and Canada that are particularly central in the collaboration network, is the least internationally oriented continent, with an average value of international share equal to 0.14, and the lowest standard deviation (0.16) of the sample. Besides Oceania, which only records two countries and therefore does not allow to build sufficiently meaningful reasoning, Africa is the most internationally oriented continent (0.36). Data show how, in general, research groups, departments, and laboratories based in Africa are seeking to engage in networks of collaboration across borders and how, instead, more central countries in terms of scientific production show a lower tendency to work with foreign entities. This finding relates to the debate regarding the necessity of favoring inclusion and diversity in the broader AI research landscape considering that African countries are, on average, the least central both in binary and weighted centrality scores and show the lowest values of relative presence overall (0.0042). Conversely, American countries are on average the most present but, as reported above, show the lowest values in terms of international collaborations. When analyzing data by discriminating per continent, patterns unfold that indicate how peripheral countries struggle in engaging with more central countries. This is, along with the structural disconnections in the authorship network, another fundamental obstacle in the formation of a global community of scholars and institutions working at the intersection between AI and crime-related research problems. 
\begin{table}[!htb]
\footnotesize
\centering
\caption{Average Values and Standard Deviation (Between Parentheses) of Binary Centrality, Weighted Centrality, Relative Presence and International Share per Continent. Note: Oceania only includes Australia and New Zealand.}
\begin{tabular}{lcccc}
\hline
\textbf{Continent} & \multicolumn{1}{c}{\textbf{Binary Centr.}} & \multicolumn{1}{c}{\textbf{Weight. Centr.}} & \multicolumn{1}{c}{\textbf{Relative Pres.}} & \multicolumn{1}{c}{\textbf{Int. Share}} \\\hline\hline
Africa & 0.0346 (0.0217) & 0.0001 (0.0001) & 0.0042 (0.2362) & 0.3588 (0.0035) \\
America & 0.1103 (0.1335) & 0.0021 (0.0049) & 0.0441 (0.0049) & 0.1408 (0.0905) \\
Asia & 0.0584 (0.0512) & 0.0006 (0.0011) & 0.0185 (0.0011) & 0.2428 (0.0321) \\
Europe & 0.0573 (0.0579) & 0.0003 (0.0005) & 0.0106 (0.0005) & 0.2751 (0.0146) \\
Oceania & 0.0909 (0.0735) & 0.0007 (0.0009) & 0.0210 (0.0009) & 0.4271 (0.0255)\\\hline
\end{tabular}

\label{continent}
\end{table}

\section{Where To, Now? Discussion and Future Developments}
The exponential diffusion of AI applications in many scientific domains outside of traditional areas in which intelligent algorithms are developed, like computer science, engineering, and mathematics, has influenced also research on criminal behavior and crime-related topics. This process has been favored by several factors. These include the increasing open availability of data on crimes and offenders, the interest of scholars from disciplines outside of criminology and social sciences for such topics and the growing accessibility of AI algorithms via statistical software and programming languages. 

In spite of these aspects, research lacks an assessment of published works at the intersection between AI and crime. The present work attempts to fill this gap providing a quantitative analysis of literature in this area. Data are gathered from Scopus, an electronic database containing over 69 million records, and are analyzed using network science. The performed search provided a total of 692 research items, temporally distributed from 1981 to 2020. 

The analysis is divided into two main dimensions. First, keyword co-occurrence graphs are investigated, using both authors- and index-keywords, to highlight patterns of themes and topics in the literature. Data indicate that scientists publishing in this area are mostly interested in cyber-related criminal topics such as \textit{Cyber-crime}, \textit{Malware}, \textit{Phishing}, and \textit{Intrusion Detection}. Conversely, topics that have gained the attention of non-specialists, activists, and policy-makers after several scandals such as algorithmic fairness, discrimination, bias, and transparency are largely overlooked. Furthermore, the analysis indicates that the higher the centrality of a keyword, the higher the number of citations that a work using that keyword will receive. 

Second, co-authorship and country-level collaboration networks are considered to assess the structure of scientific collaboration at the individual and national levels. The graph of author-collaboration reveals a highly disconnected structure: the total 1,964 scientists that have authored at least one work in the sample are divided into many components that include 59 isolates, 134 dyads, and 115 triads. When countries are taken into account, considering the primary affiliation of authors (to exemplify: if a research $A$ publishes a paper while affiliated with Harvard University, its country-affiliation will be processed as \textit{United States}), further patterns emerge. Most central countries (countries that have a higher number of international collaborations without considering domestic ones) and most prevalent ones (namely countries that are more present than the others when affiliations are considered) tend to be less internationally collaborative when controlling for feedback loops. This means that, on average, researchers from these countries (e.g., United States, China, India) prefer to collaborate with scientists affiliated with institutions based in the same country. 

These two layers of findings can help in shaping broader discussions regarding the interplay between the current state of research at the intersection between AI and crime and its future developments. Given that Scopus data show that works in this area are, in proportion, increasing more in terms of quantity compared to works covering AI problems overall, it is crucial to assess the likely pathways that this research area may take tomorrow. With regard to themes and topics, the large interest in cyber-related themes suggests, in contrast, the underdevelopment of applications in other relevant criminological or crime-related areas. Additionally, analyses reveal how scientists are overlooking critical topics regarding ethics and responsible use of intelligent algorithms in areas such as criminal justice and policing. Given that keyword, centrality is tightly related to citations, and that citations can predict future research directions, resource allocations and even recruitment processes \citep{BurrellPredictingfuturecitation2003, ErdiPredictionemergingtechnologies2013, BaiPredictingcitationsscholarly2019}, it is necessary to timely increase the number of works focusing on ethics and related matters to enhance scientific debate on the need for responsible use of algorithms. Responsible use of algorithms encompasses several issues, such as avoidance of machine-bias and discrimination against minorities or disadvantaged strata of the population. Given that algorithmic decision-making is increasingly deployed in the real world, impacting the life of millions of citizens worldwide, the attention on technical applications of AI systems in crime-related problems should be balanced with works that focus on societal, political, legal and moral consequences of such intelligent systems.

For what concerns co-authorship and country-level collaboration patterns, additional considerations cane made. First of all, the highly disconnected structure of co-authorship may represent an obstacle to the development of a structured and solid community. Given the highly trans-disciplinary nature of the debate at the intersection between AI and crime, scientific collaboration is crucial to guarantee a debate that overcomes structural barriers and asymmetries. In fact, if researchers will continue to publish within this component-based structure, it will become difficult to establish grounded debates and inclusive cooperation. Inclusive cooperation indeed represents an issue, given the current state of international collaboration. Due to several causes (e.g., the disparity of resources, critical domain of application), most central and productive countries tend to avoid international collaborations. Contrarily, developing countries that are, in general, less productive, are trying to engage in international partnerships to counterbalance the lower availability of funds, grants, and resources to conduct research in this area. This asymmetry reinforces exclusion in international research and disallows such peripheral countries in joining scientific production and debate. A Western-centric standpoint in discussing applications and consequences of AI systems applied to investigate or reduce crime-related problems can reinvigorate structural differences between countries. Given that research in this area often refers to the possibility to deploy intelligent systems in real-world scenarios, it is fundamental to avoid the risks of such future increasing inequalities. Less disconnected and more transnationally-oriented scientific collaboration both at individual and country levels can help in addressing these aspects.
\newpage

\bibliography{review_AIC.bib}
\end{document}